\documentclass{ametsocV6.2}
\usepackage{mathtools}
\usepackage{booktabs}
\usepackage{array}
\newcolumntype{L}[1]{>{\raggedright\arraybackslash}p{#1}}
\usepackage{hyperref}

\nolinenumbers                   
\let\internallinenumbers\relax   
\usepackage[font=small,labelsep=period]{caption}

\title{Su\^{e}tes: An end-to-end differentiable non-hydrostatic limited-area dynamical core}

\authors{Tim Whittaker\aff{a,b,*}\correspondingauthor{Tim Whittaker, tim.whittaker@ubc.ca}, Seth Taylor\aff{c}, Elsa Cardoso-Bihlo\aff{d}, Alejandro Di Luca\aff{b}, Alex Bihlo\aff{d,*}}

\affiliation{\aff{a}{Department of Earth, Ocean and Atmospheric Sciences, University of British Columbia, Vancouver, BC, Canada}\\
\aff{b}{Centre pour l'étude et la simulation du climat à l'échelle régionale (ESCER), D\'epartement des Sciences de la Terre et de l’Atmosph\`ere, Universit\'e du Qu\'ebec \`a Montr\'eal, Montr\'eal,  Qu\'ebec, Canada}\\
\aff{c}{Department of Computer Science, University of Saskatchewan, Saskatoon, Canada}\\
\aff{d}{Department of Mathematics and Statistics, Memorial University of Newfoundland, St. John's, Canada}\\
\aff{*}{Equal Contributions}}

\abstract{We present Suêtes, a fully differentiable, non-hydrostatic, limited-area atmospheric dynamical core implemented in JAX. Suêtes includes both a semi-implicit semi-Lagrangian scheme and a Split-Explicit Runge–Kutta scheme with acoustic substepping, enabling applications from coarse regional simulations to convection-permitting dynamics. Its terrain-following geometry, spatial discretization, time integration, initial conditions, lateral boundary treatment, and included physical parameterizations are all compatible with reverse-mode automatic differentiation. A single reverse-mode pass can therefore compute sensitivities of forecast diagnostics to selected initial and boundary fields, physical parameters, topography, and computational geometry. We provide a series of numerical experiments---including a tracer-source inverse problem, a sensitivity analysis of a three-dimensional squall-line, adversarial perturbation construction an ERA5-driven downslope hurricane force wind event, topography optimization, and terrain-following coordinate optimization---demonstrating the versatility of Suêtes as a framework for gradient-based diagnosis and model development at regional and convection-permitting scale. }

\begin{document}

\maketitle

\section{Introduction}
Limited-area atmospheric models have become essential tools for the study of weather and climate processes and phenomena, since they can resolve such fine-scale detail at a fraction of the computational cost of global models. These models are currently run operationally at kilometer scales and are used for a variety of important tasks, including numerical weather prediction (NWP, \cite{Bauer2015,mctaggart-cowan_modernization_2019}), regional climate modeling \citep{laprise_regional_2008,jones_coordinated_2011,giorgi_regional_2015,Denis2002, DiLuca2015, Prein2015}, and the study of local, fine-scale processes such as deep convection \citep{Rotunno1988, Houze2004, Schumacher2020}, orographic precipitation and terrain-forced convection \citep{Roe2005, Houze2012, Kirshbaum2018}, and severe downslope windstorms \citep{KlempLilly1975, Smith1985, Durran1990, Stewart01061995}. These limited-area models have been under continuous development for decades, among them the Weather Research and Forecasting (WRF) model \citep{skamarock2008time}, the Canadian MC2 and its successor the Global Environmental Multiscale (GEM) model  \citep{TheCanadianMC2ASemiLagrangianSemiImplicitWidebandAtmosphericModelSuitedforFinescaleProcessStudiesandSimulation,cote_operational_1998}, the AROME-France model based on the Aladin-NH dynamical core \citep{TheAROMEFranceConvectiveScaleOperationalModel,https://doi.org/10.1002/qj.522}, the COSMO model \citep{klasa_evaluation_2018,steger_regional_2020} and its successor the ICON-CLM model \citep{pham_icon_2021,egusphere-2025-4726}. All these models can run at convection-permitting scales (i.e., with horizontal grid spacings below about 4 km) thanks to their non-hydrostatic formulation of the equations governing atmospheric motion \citep{Prein2015}.

Alongside these developments, recent advances in machine learning (ML) have begun to reshape atmospheric modelling, opening new approaches to representing and predicting atmospheric processes. Purely data-driven global models can now attain skill comparable to NWP operational systems at a fraction of the inference cost \citep{Bi2023, Lam2023, https://doi.org/10.1029/2023MS004019}. However, they struggle to generalize to long-duration simulations and statistically rare weather events such as those related to extreme weather \citep{Pasche2025, Zhang2025}. A new class of hybrid models sits between these ML approaches and conventional physics-based systems and has also shown competitive skill at reduced computational cost while retaining essential physical constraints \citep{Kochkov2024, pereira2026learningadvectneuralsemilagrangian}. For instance, NeuralGCM utilizes a differentiable spectral dynamical core coupled to a neural network parameterization of subgrid-scale physical processes, which is trained online \citep{Kochkov2024}. This training paradigm is made possible in large part because the underlying code is written in a differentiable programming framework, allowing gradients to be back-propagated through the entire model and used for gradient-based optimization. Differentiable models have since appeared at intermediate complexity \citep{egusphere-2025-6266, https://doi.org/10.1029/2025MS005615}, in large-eddy simulation \citep{gmd-19-1103-2026}, and in hybrid formulations where trained networks replace individual numerical operators, correct the discrete solution in the loop, or make fixed parameters state-dependent \citep{Wang2026, pereira2026learningadvectneuralsemilagrangian, https://doi.org/10.1029/2025MS004969, nath2026makingtunableparametersstatedependent}. The same gradients can also be used for diagnostic purposes, including to construct extreme-event storylines \citep{Whittaker2026, whittaker2025constructingextremeheatwavestorylines, hakim2026grayswanfactorymaking}, analyze model sensitivities \citep{Bano-Medina2025}, and probe the limits of predictability \citep{AtmosphericPredictabilityBeyond30DayswithMachineLearning}.

To our knowledge, no differentiable limited-area dynamical core currently exists; all of the above examples operate at the global scale. Purely data-driven methods exist at the regional scale, such as diffusion models trained to map a coarse global state onto a kilometre-scale regional one. These have been used in producing 2\,km fields over Taiwan from a 25\,km reanalysis \citep{Mardani2025} and in emulating convection-allowing forecasts directly \citep{pathak2024kilometerscaleconvectionallowingmodel, perkins2025hiroacefastskillfulai}. Another example includes globally stretched grid refined over the area of interest and trained on a mix of RCM and ERA5 data \citep{Bano-Medina2025a, Regionaldatadrivenweathermodelingwithaglobalstretchedgrid}. However, since there are no limited-area differentiable dynamical cores, the techniques of hybrid modelling in the regional context have not seen such development. Without gradients through the dynamics, a learned component cannot be optimized over a model rollout, and the model sensitivities that support gradient-based diagnosis are equally unavailable. 

In this work, we present Su\^etes, a non-hydrostatic limited-area dynamical core written entirely in JAX, integrating the fully compressible Euler equations on a staggered Arakawa C-grid \citep{Arakawa1977} under a general terrain-following vertical coordinate, with lateral relaxation toward driving data \citep{davies1976lateral}, flux-form tracer transport \citep{lin1996multidimensional}, and a compact differentiable physics package. The main purpose of Suêtes is to make high-resolution regional atmospheric simulations differentiable, so that gradients can be used for optimization, inverse problems, and machine-learning applications. The model is named after \emph{les su\^etes}, the southeasterly downslope windstorms that descend the western escarpment of the Cape Breton Highlands, whose forcing ridge is only tens of kilometres wide and vanishes from the smoothed orography of a global model \citep{mcildoon2008suetes, Stewart01061995}. Three properties distinguish Su\^etes from the differentiable atmospheric models described above. First, a limited-area domain reaches convection-permitting grid spacings, whereas differentiable global cores have so far been run at $\mathcal{O}(100)$\, km or coarser \citep{Kochkov2024, egusphere-2025-6266}. Since the cost of a training rollout scales with the number of degrees of freedom carried through the reverse pass, a regional domain makes online optimization affordable at resolutions that would be prohibitive globally. Second, the hydrostatic approximation underlying existing differentiable atmospheric cores breaks down at these scales, and Su\^etes retains the vertical acceleration and acoustic modes required by a kilometre-scale solution. Third, gradients propagate through time integrators of a complexity not previously differentiated in this setting, comprising a two-time-level semi-implicit semi-Lagrangian scheme in the lineage of the Canadian MC2 model \citep{thomas1998a}, whose coupled acoustic-gravity system is inverted by preconditioned GMRES \citep{saad1986gmres}, and a Split-Explicit Runge--Kutta scheme with acoustic subcycling following the WRF model \citep{klemp2007conservative, wicker2002time}.

The remainder of the paper is organised as follows. Section~\ref{sec:DynamicalCore} presents the governing equations, coordinate transformations and metric operators, C-grid discretization, both time integrators, boundary treatment, and basic physics package. Section~\ref{Sec:Results} presents several forward benchmarks together with their associated inverse and sensitivity demonstrations. Conclusions follow in Section~\ref{sec:conclusions}. Numerical convergence and a compact hardware-scaling experiment are documented in Appendix~\ref{app:convergence_verification}.

\section{Su\^{e}tes' dual dynamical cores}\label{sec:DynamicalCore}

Su\^{e}tes features two distinct, runtime-selectable dynamical cores. The first is a semi-implicit semi-Lagrangian (SISL) scheme inspired by the early Canadian MC2 model \citep{TheCanadianMC2ASemiLagrangianSemiImplicitWidebandAtmosphericModelSuitedforFinescaleProcessStudiesandSimulation} and used across a number of current atmospheric models, including the European Centre for Medium-Range Weather Forecasts (ECMWF) Integrated Forecasting System \citep[IFS; ][]{temperton_twotimelevel_2001}, the Canadian Global Environmental Multiscale model \citep[GEM; ][]{cote_operational_1998}, and the Met Office Unified Model \citep[UM; ][]{wood_inherently_2014}. Trajectory-based advection circumvents the advective CFL stability constraint, while implicit treatment of the acoustic and gravity-wave terms alleviates the fast-mode restriction, so time steps far exceeding those of an explicit formulation become admissible. This advantage is greatest at grid spacings of roughly 10~km and coarser, where an explicit time step would be set by stability rather than by accuracy. The second core is a strictly Eulerian, Split-Explicit Runge--Kutta scheme with acoustic subcycling \citep{skamarock2008time, klemp2007conservative} intended for convection-permitting configurations (horizontal resolution $\lesssim 4$~km). At these scales, the SISL efficiency margin narrows, and the Split-Explicit core becomes more accurate \citep{Steppeler2003}. This core also avoids the interpolation inherent to semi-Lagrangian trajectory calculations, which damps the short wavelengths carrying much of the resolved convective signal. Each time-stepping scheme shares a common finite-difference spatial discretization in the computational domain defined by an oblique stereographic projection and a general terrain-following vertical coordinate. In this section, we describe these components of Suêtes in detail, along with the treatment of boundary conditions, the solver stabilization, and further details on the numerical implementation.

\subsection{Horizontal domain and vertical coordinate system}

Let $(x,y) \in \Omega_h$ denote the horizontal coordinates for the limited area horizontal domain $\Omega_h$, defined by an oblique stereographic projection tangent to the Earth at the central latitude and longitude $(\phi_c, \lambda_c)$ of the model domain. The computational coordinates are $(x,y, \zeta) \in \hat\Omega \coloneqq \Omega_h \times [0, L_z]$. The physical domain is obtained from a terrain-following coordinate transformation $(x,y, \zeta) \mapsto (x,y,z)$ defined by
\begin{equation}
\label{eq:verticalCoordinate}
z = Z(x, y, \zeta),\quad  \text{with} \quad  Z(x,y,0) = h_0(x,y) \,, \quad Z(x,y,L_z) = L_z \,, \quad \frac{\partial Z}{\partial \zeta } > 0 \,,
\end{equation}
where $h_0(x,y)$ is the bottom topography. Su\^{e}tes includes four vertical coordinate transforms, all implemented as differentiable routines: the classical Gal-Chen Somerville coordinate \citep{GALCHEN1975209}, a hybrid height coordinate with exponential decay of terrain influence, the full SLEVE coordinate \citep{ANewTerrainFollowingVerticalCoordinateFormulationforAtmosphericPredictionModels}, and the recently developed NEUVE coordinate based on an integral neural transform \citep{whittaker2025learningverticalcoordinatesautomatic}.

\subsection{Differential operators in terrain-following coordinates}\label{sec:diff_operators}

The combination of a horizontal projection plane and a terrain-following vertical coordinate introduces metric scale factors into the differential operators of the governing equations. The horizontal Cartesian coordinates $(x, y)$ are defined on the conformal projection plane where physical distances are related to projected distances through the conformal map-scale factor
\begin{equation*}
m(x, y) = 1 + \frac{x^2 + y^2}{4 R_e^2} \,,
\end{equation*}
where $R_e$ is the radius of the Earth. The inverse projection yields the geographic latitude $\phi(x,y)$, from which the Coriolis parameter is determined as $f(x, y) = 2 \Omega_e \sin \phi(x,y)$, where $\Omega_e$ is the Earth's angular velocity.

By applying the chain rule of differentiation from the physical frame $(x/m, y/m, z)$ to the computational frame $(x, y, \zeta)$, the spatial gradient of any scalar field $\chi=\chi(x, y, \zeta, t)$ along the physical orthonormal directions $(\boldsymbol{e}_x, \boldsymbol{e}_y, \boldsymbol{e}_z)$ at fixed time is given by
\begin{equation}
\tilde\nabla \chi = m \left( \frac{\partial \chi}{\partial x} - \frac{Z_x}{Z_\zeta} \frac{\partial \chi}{\partial \zeta} \right) \boldsymbol{e}_x + m \left( \frac{\partial \chi}{\partial y} - \frac{Z_y}{Z_\zeta} \frac{\partial \chi}{\partial \zeta} \right) \boldsymbol{e}_y + \frac{1}{Z_\zeta} \frac{\partial \chi}{\partial \zeta} \boldsymbol{e}_z \,,
\label{eq:grad_transform}
\end{equation}
where $Z_{x} = \partial_x Z$, $Z_y = \partial_y Z$, and $Z_\zeta = \partial_\zeta Z$ is the vertical metric layer thickness.

The physical velocity field in local orthogonal directions will be denoted by $\boldsymbol{U} = (u, v, w)$, where $(u, v)$ represent the physical horizontal velocity components along the conformal projection axes, and $w = \dot{z}$ is the physical vertical velocity. The corresponding contravariant (computational) velocities $(\dot{x}, \dot{y}, \dot{\zeta})$ representing the rates of change across grid coordinates are 
\begin{equation}
\dot{x} = m u \,, \qquad \dot{y} = m v \,, \qquad \dot{\zeta} = \frac{1}{Z_\zeta} \left( w - m Z_x u - m Z_y v \right) \,.
\label{eq:contravariant_vel}
\end{equation}
The contravariant vertical velocity $\dot{\zeta}$ represents the rate at which fluid parcels cross the terrain-following $\zeta$-surfaces. The material derivative of $\chi$ can therefore be expressed as
\begin{equation}
\frac{D \chi}{D t}
= \frac{\partial \chi}{\partial t} + \boldsymbol{U}\cdot\tilde\nabla \chi
= \frac{\partial \chi}{\partial t}
  + m u\frac{\partial \chi}{\partial x}
  + m v\frac{\partial \chi}{\partial y}
  + \dot{\zeta}\frac{\partial \chi}{\partial\zeta} \,.
\label{eq:material_transform}
\end{equation}
This operator is applied component-wise to the prognostic fields. The geometric rotation of the horizontal basis is represented separately by the curvature term in the momentum equation below. At the lower and upper boundaries where $\dot{\zeta} = 0$, Eqn.~\eqref{eq:contravariant_vel} naturally enforces the kinematic flow boundary conditions
\begin{equation*}
w(x,y,0) = m(u Z_x + v Z_y) \,, \qquad w(x,y,L_z) = 0 \,,
\end{equation*}
along the terrain and at the rigid top lid, respectively. Using the contravariant coordinate components, the physical divergence of any vector field $\boldsymbol{V} = (V^x, V^y, V^z)$ can be written in the conservative form
\begin{align}
\begin{split}
\tilde\nabla \cdot \boldsymbol{V} &= \frac{1}{J} \left[ \frac{\partial}{\partial x} \left( J m V^x \right) + \frac{\partial}{\partial y} \left( J m V^y \right) + \frac{\partial}{\partial \zeta} \left( J V^\zeta\right) \right]\\
&= \frac{m^2}{Z_\zeta} \left[ \frac{\partial}{\partial x} \left( \frac{V^x Z_\zeta}{m} \right) + \frac{\partial}{\partial y} \left( \frac{V^y Z_\zeta}{m} \right) + \frac{1}{m^2} \frac{\partial}{\partial \zeta} \left(V^z - m Z_x V^x - m Z_y V^y\right) \right] \,,
\label{eq:div_transform}
\end{split}
\end{align}
where $J = Z_\zeta / m^2$ is the physical volume element per computational coordinate unit and the contravariant vertical component associated with $\boldsymbol{V}$ is
\begin{equation*}
V^\zeta
=
\frac{1}{Z_\zeta}
\left(V^z-mZ_xV^x-mZ_yV^y\right).
\end{equation*}

\subsection{Governing equations}

The governing equations solved in both dynamical cores are the Euler equations with the prognostic variables $(\boldsymbol{U}, \pi, \theta_v)$, where $\pi = (p/p_0)^{R_d/c_p}$ is the Exner pressure and $\theta_v$ is the virtual potential temperature. Following the formulations used in the WRF model \citep{skamarock2008time} and the Model for Prediction Across Scales (MPAS) \citep{skamarock2012multiscale}, and to isolate the fast linear acoustic and gravity wave modes for semi-implicit and time-split integration, the thermodynamic fields are decomposed into a horizontally homogeneous, time-invariant hydrostatic reference state and a prognostic perturbation
\begin{equation*}
\pi(x, y, \zeta, t) = \bar{\pi}(z) + \pi'(x, y, \zeta, t) \,, \qquad 
\theta_v(x, y, \zeta, t) = \bar{\theta}_v(z) + \theta_v'(x, y, \zeta, t) \,,
\end{equation*}
where the reference profile exactly satisfies the discrete hydrostatic relation $\partial \bar{\pi}/\partial z = -g/(c_p \bar{\theta}_v)$. The reference state is constructed at model initialization either analytically from a prescribed constant Brunt--V\"ais\"al\"a frequency $N$ for idealized benchmarks, or from the horizontal mean of the driving ERA5 atmospheric column and interpolated onto the model levels. Using these thermodynamic variables, the mass density becomes diagnostic, satisfying
\begin{equation}
\rho = \frac{p_0}{R_d\, \theta_v}\, \pi^{\,c_v/R_d} \,.
\label{eq:eos}
\end{equation}
Using the curvilinear material derivative~\eqref{eq:material_transform}, spatial gradients~\eqref{eq:grad_transform}, and divergence~\eqref{eq:div_transform}, the governing equations on the terrain-following grid $(x, y, \zeta)$ are given by
\begin{subequations}\label{eq:governing_eqns}
\begin{align}
\frac{D\boldsymbol{U}}{D t}  + c_p\, \theta_v \tilde\nabla \pi' + (f + M_c) \boldsymbol{e}_z \times \boldsymbol{U} &= g \frac{\theta_v'}{\bar{\theta}_v} \boldsymbol{e}_z + \boldsymbol{D} + \boldsymbol{P} \,, \label{eq:U}\\[2pt]
\frac{D \pi'}{D t} + C_\pi \tilde\nabla \cdot (\bar{\rho}\, \bar{\theta}_v\, \boldsymbol{U}) &=0  \,, \label{eq:pi}\\[2pt]
\frac{\partial \theta_v}{\partial t} + \boldsymbol{U} \cdot \tilde\nabla \theta_v &= D_{\theta_v} + P_{\theta_v} \,, \label{eq:thv}\\[2pt]
\frac{\partial (\rho q_i)}{\partial t} + \tilde\nabla \cdot (\rho q_i \boldsymbol{U}) &= \rho M_i \,, \label{eq:adv}
\end{align}
\end{subequations}
where $C_\pi = R_d \bar{\pi}/(c_v \bar{\rho}\, \bar{\theta}_v)$ is the acoustic compressibility coefficient, $f(x,y)$ is the Coriolis parameter, and $M_c = v \partial_x m  - u \partial_y m = (x v - y u)/2 R_e^2$ represents the geometric curvature metric acceleration arising from spatial gradients of the conformal map scale factor. The terms $\boldsymbol{D}$ and $D_{\theta_v}$ denote optional core-level diffusion and stabilization applied to momentum and virtual potential temperature (cf.~Sect.~\ref{sec:diffusion}). They are distinguished from $\boldsymbol{P}$ and $P_{\theta_v}$, which denote tendencies supplied by physical parameterizations through the differentiable physics interface (cf.~Sect.~\ref{sec:physics}). This distinction concerns their role in the model in that a state-dependent turbulent closure is included in $\boldsymbol{P}$ even if its mathematical form is diffusive. The specific mass of tracer or moisture species $i$ is given by $q_i$, and $M_i$ denotes its net non-advective source or sink per unit mass of air, including any contributions from physical parameterizations or prescribed emissions. In moist simulations, specific masses $q_i$ (such as water vapour, cloud water, and hydrometeor condensates) are advected strictly in conservative flux form \eqref{eq:adv} using a Flux-Form Semi-Lagrangian (FFSL) scheme applied to the partial densities $\rho q_i$. Following advection and lateral boundary blending relative to the driving data (generally reanalysis or a coarser resolution simulation), the diagnostic relation \eqref{eq:eos} is re-applied at the end of each time step to reconcile density with the updated potential temperature and pressure fields. The formulation of the continuity equation \eqref{eq:pi} in terms of Exner pressure perturbation $\pi'$ with static background density weighting ($\bar{\rho} \bar{\theta}_v$) ensures that the acoustic matrix operator remains strictly linear when the background profile is held fixed. This property is essential for the efficient convergence of the Generalized Minimal Residual (GMRES) Krylov solver within the semi-implicit semi-Lagrangian dynamical core.

\subsection{Spatial discretization}
\label{sec:spatial_discretization}

\begin{figure}
    \centering
    \includegraphics[width=\linewidth]{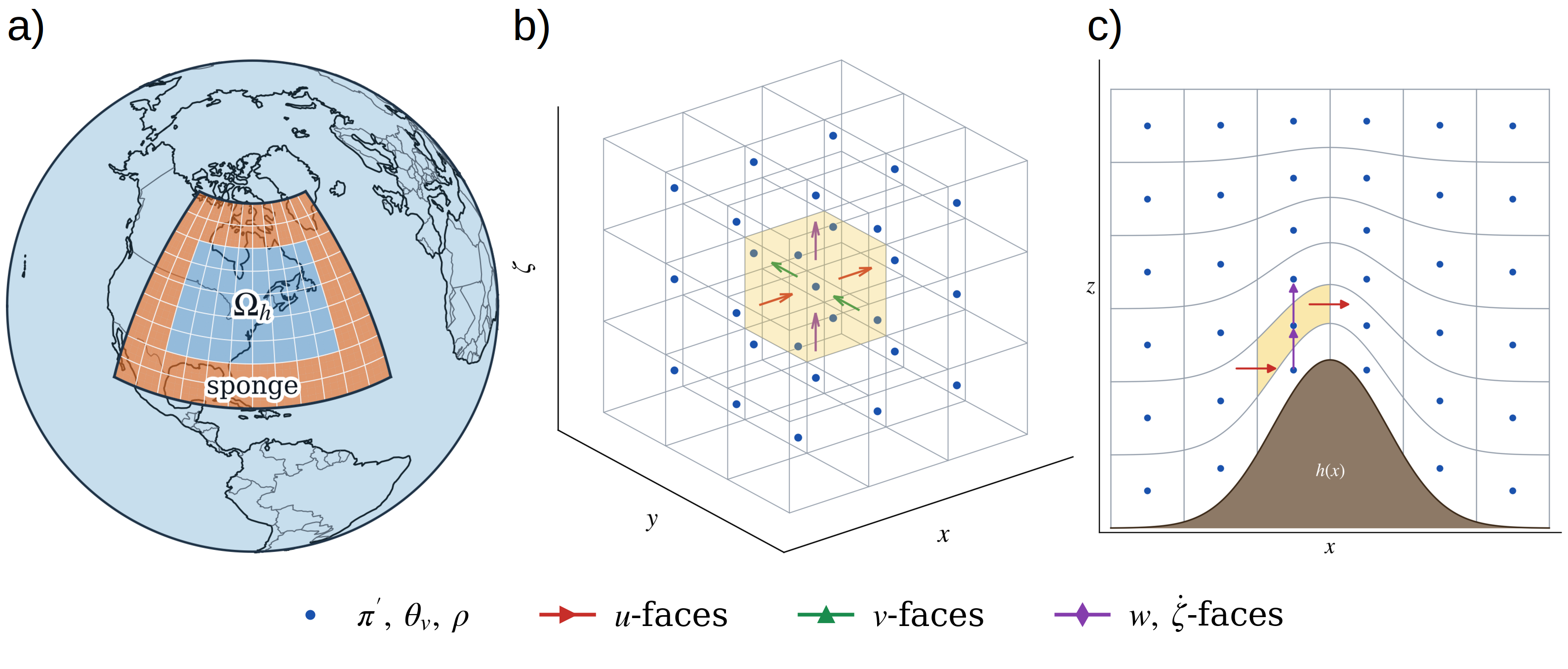}
    \caption{Su{\^e}tes Spatial Discretization. A stereographic projection of the limited-area domain defines the computational domain with a metric induced by the terrain-following coordinate in the physical domain. }
    \label{fig:spatial_discretization}
\end{figure}

Su\^{e}tes' computational grid employs an Arakawa C-grid \citep{Arakawa1977} horizontally and a Lorenz grid~\citep{Lorenz1960} vertically across both dynamical cores (see Fig.~\ref{fig:spatial_discretization}). The thermodynamic variables $\pi'$, $\theta_v$, and $\rho$, together with all moisture and chemical tracer fields $q_i$, are discretized at cell centres. The $u$- and $v$-faces carry the normal horizontal velocity components, while the $w$-faces carry both the physical vertical velocity $w$ and the contravariant vertical velocity $\dot{\zeta}$. In the grid interior, horizontal derivatives are approximated by second-order centred differences between adjacent staggered locations. For example, the mass-point-to-$u$-face gradient and $u$-face-to-mass-point flux divergence are
\begin{equation*}
\left(\delta_x a\right)_{i+\frac12,j,k}
=
m_{i+\frac12,j}
\frac{a_{i+1,j,k}-a_{i,j,k}}{\Delta x},
\qquad
\left(\delta_x F^x\right)_{i,j,k}
=
m_{i,j}
\frac{F^x_{i+\frac12,j,k}-F^x_{i-\frac12,j,k}}{\Delta x},
\label{eq:discrete_horizontal_derivatives}
\end{equation*}
with analogous expressions in the $y$ direction. Quantities required at different staggered grid locations are computed using two-point arithmetic averaging rules such as,
\begin{equation*}
\left(\mathcal{A}_x a\right)_{i+\frac12,j,k}
=
\frac{1}{2}\left(a_{i+1,j,k}+a_{i,j,k}\right).
\label{eq:discrete_staggered_average}
\end{equation*}
The discretized operators $\tilde{\nabla}_h$ and $\tilde{\nabla}_h\!\cdot$ are then obtained by substituting the staggered differences and averages into the terrain-following gradient and conservative flux-divergence expressions \eqref{eq:grad_transform} and \eqref{eq:div_transform}, respectively. Map factors and vertical-coordinate metrics are then evaluated, at, or arithmetically averaged to the staggered location of the resulting variable, respectively.

\subsection{Temporal discretization}

After spatial discretization, we collect all degrees of freedom into a single state vector $\boldsymbol{\Psi}_h = \left(\boldsymbol{U}_h, \pi'_h, \theta'_{v,h}\right)$, where the subscript~$h$ denotes the spatial discretization of the prognostic variables on the mesh. We can then write the semi-discretized governing equations in the form 
\begin{equation}
\label{eq:semi_discretized_governing_eqs}
\frac{\partial \boldsymbol{\Psi}_h}{\partial t} = \mathcal{L}\boldsymbol{\Psi}_h + \mathcal{A}(\boldsymbol{\Psi}_h) + \mathcal{R}(\boldsymbol{\Psi}_h) \coloneqq \mathcal{L}\boldsymbol{\Psi}_h + \mathcal{N}(\boldsymbol{\Psi}_h)\,,
\end{equation}
where $\mathcal{L}$ contains the linear acoustic and gravity-wave coupling about the hydrostatic reference state
\begin{equation}
\mathcal{L}\boldsymbol{\Psi}_h = \begin{pmatrix}
-c_p \bar\theta_v \tilde\nabla_h \pi'_h + g \frac{\theta'_{v,h}}{\bar\theta_v}\boldsymbol{e}_z 
\\
- C_\pi \tilde\nabla_h \cdot \left(\bar\rho \bar\theta_v \boldsymbol{U}_h\right)
\\
- w_h \partial_z \bar \theta_v
\end{pmatrix},
\end{equation}
and where $\tilde\nabla_h$ and  $\tilde\nabla_h \cdot$ denote the finite-difference approximations of the gradient and divergence operators, respectively. The operator $\mathcal{A}$ contains the nonlinear advective terms, while $\mathcal{R}$ contains the additional nonlinearities that include the pressure-gradient, Coriolis, diffusions, and parameterized physical processes. \par

The two dynamical cores differ in their treatment of the semi-discretization \eqref{eq:semi_discretized_governing_eqs}. The SISL scheme approximates the material derivatives $\mathcal{A}$ through parcel trajectories and treats $\mathcal{L}$ implicitly over the full time step. The Split-Explicit Runge--Kutta core advances the nonlinearity $\mathcal{N}$ using a large Eulerian time step and performs a smaller scale integration of the linear terms. In both dynamical cores, the transport equations \eqref{eq:adv} are solved using a directional-split Flux-Form Semi-Lagrangian (FFSL) scheme \citep{lin1996multidimensional}. This scheme integrates the one-dimensional cumulative mass function across cell boundaries at continuous departure locations along each coordinate axis ($x, y, z$), preserving volumetric mass $\rho q_i$ exactly. This scheme benefits from avoiding the use of non-differentiable point clipping or Eulerian flux limiting operations. \par

\subsubsection{Semi-implicit semi-Lagrangian scheme}

The SISL core uses a two-time-level, off-centred discretization to compute the evolution of the prognostic fields $\chi \in \{u, v, w, \theta_v'\}$. The time-discretization can be written as
\begin{equation}
\frac{\chi^{n+1} - \chi^n_{\rm d}}{\Delta t} = \alpha\left[\mathcal{L}\boldsymbol{\Psi}_h^{n+1}\right]_\chi + (1-\alpha)\left[\mathcal{L}\boldsymbol{\Psi}_h^n\right]_{\chi,\rm d}+\left[\mathcal{R}^{n+1/2}\right]_{\chi,\rm d},
\label{eq:sisl}
\end{equation}
where $\chi^n_{\rm d}$ denotes the field value at time $t_n$ evaluated at the departure point $\mathbf{x}_{\rm d}$ of the trajectory arriving at the grid point $\mathbf{x}$ at time $t_{n+1}$. The values in the $\mathcal{R}$ term are extrapolated to the midpoint $t^{n+1/2}$ via
\begin{equation*} 
\mathcal{R}^{n+1/2} = \frac{3}{2}\,\mathcal{R}^n - \frac{1}{2}\,\mathcal{R}^{n-1} \,,
\end{equation*} 
and then evaluated at the departure point. The off-centering parameter $\alpha \in [\tfrac{1}{2}, 1)$ controls the amount of implicit damping. Departure-point evaluations of momentum $(u, v, w)$ and virtual potential temperature perturbation $\theta_v'$ are obtained using tensor-product tricubic interpolation across the 64 surrounding grid points. Because the Exner pressure perturbation $\pi'$ evolves as an Eulerian field governed directly by mass divergence and acoustic wave propagation in the continuity equation, it is not advected via semi-Lagrangian trajectories, and remains an Eulerian field update through its coupling to the discrete velocity divergence.\par

Departure points are obtained by two iterations of the midpoint trajectory equation,
\begin{equation}\label{eq:traj}
\mathbf{x}_{\rm d}^{(k+1)} = \mathbf{x} - \Delta t\, \mathbf{U}^{n+1/2}\left( \tfrac{1}{2}\big(\mathbf{x} + \mathbf{x}_{\rm d}^{(k)}\big) \right) \,,  
\end{equation} 
where $\mathbf{U}^{n+1/2} = \tfrac{3}{2}\mathbf{U}^n - \tfrac{1}{2}\mathbf{U}^{n-1}$ is the midpoint-extrapolated velocity field, with trilinear interpolation used inside the iteration and tricubic interpolation reserved for the final field evaluation. During automatic differentiation or adjoint evaluations, gradients flow only through the final interpolation weights but not through the iterative refinement of $\mathbf{x}_{\rm d}$.

To absorb upward-propagating gravity waves before they reflect from the rigid lid, a Rayleigh sponge profile is applied to $w$ above an altitude $z_{\rm sponge}$,
\begin{equation}
\tau_{\rm sponge}(z) = \begin{cases} 0, & z \leqslant z_{\rm sponge} \,, \\[2pt] \tau_{\max}\, \tfrac{1}{2}\!\left[1 + \tanh\!\left(\pi\,\dfrac{z - z_{\rm sponge}}{L_z - z_{\rm sponge}} - \dfrac{\pi}{2}\right)\right], & z > z_{\rm sponge} \,. \end{cases}
\label{eq:rayleigh}
\end{equation}
This damping enters the vertical momentum equation as an additive $-\tau_{\rm sponge}\, w$ on the left-hand side, which is folded directly into the implicit operator.

After explicit evaluations and semi-Lagrangian interpolation yield an intermediate state $\boldsymbol{\Psi}_h^\ast$, the implicit update is obtained by solving the linear system
\begin{equation}
\big(\mathbf{I} - \alpha\,\Delta t\, \mathcal{L}\big)\,\boldsymbol{\Psi}_h^{n+1} = \boldsymbol{\Psi}_h^\ast \,.
\label{eq:implicit}
\end{equation}
This system couples $u$, $v$, $w$, $\pi'$, and $\theta'_v$, with $\dot{\zeta}$ diagnosed consistently from the updated velocity. We solve \eqref{eq:implicit} using restarted GMRES \citep{saad1986gmres}, an iterative method used extensively in non-hydrostatic atmospheric modeling \citep{giraldo2010semi, thomas1998a}. To accelerate convergence, we construct an analytical one-dimensional vertical preconditioner by formally eliminating $u$, $v$, $w$, and $\dot{\zeta}$ from \eqref{eq:implicit} under the assumption that horizontal gradients of $\pi'$ vanish \citep{precond}. This collapses the coupled system to a decoupled tridiagonal Helmholtz problem for $\pi'_k$ across each $(i, j)$ vertical column
\begin{equation*}
-Q_\pi\, P_{k-1/2}\, \pi'_{k-1} + \big(1 - Q_\pi (P_{k-1/2} + P_{k+1/2})\big)\, \pi'_k - Q_\pi\, P_{k+1/2}\, \pi'_{k+1} = \mathrm{RHS}_k \,,
\end{equation*}
with coefficients
\begin{equation*}
Q_\pi = \frac{\alpha\, \Delta t\, C_\pi}{\Delta z_m} \,, \qquad P_{k \pm 1/2} = \frac{\alpha\, \Delta t\, c_p\, \bar{\theta}_{v,k\pm 1/2}}{\Delta z_{w,k\pm 1/2}\,(1 + \Delta t\, \tau_d)}\, \bar{\rho}_{k\pm 1/2}\, \bar{\theta}_{v,k\pm 1/2} \,,
\end{equation*}
subject to homogeneous boundary conditions at the surface and rigid lid. Back-substitution recovers $w$ and $\dot{\zeta}$ in physical and contravariant form. GMRES typically converges in a small number of outer iterations when initialized with the solution from the previous time step.

\subsubsection{Split-Explicit Runge--Kutta scheme}
\label{sec:split_explicit}

For high-resolution and convection-permitting simulations where Eulerian conservation is paramount, Su\^{e}tes provides a Split-Explicit third-order Runge--Kutta (RK3) scheme based on the formulation of \cite{klemp2007conservative} and \cite{wicker2002time}. The method advances the nonlinear tendencies $\mathcal{N}(\boldsymbol{\Psi}_h)$ over the large time step while integrating the fast linear acoustic and gravity-wave terms $\mathcal{L}\boldsymbol{\Psi}_h$ using smaller acoustic substeps. Let $\mathcal{E}_T(\boldsymbol{\Psi},\boldsymbol{F})$ denote the acoustic-subcycled evolution over a stage interval $T$, starting from the state $\boldsymbol{\Psi}$ with the vector-valued slow tendency $\boldsymbol{F}$ held fixed. The RK3 stages are
\begin{align*}
&\boldsymbol{\Psi}_h^\ast
= \mathcal{E}_{\Delta t/3}
   \left(\boldsymbol{\Psi}_h^n,\,\mathcal{N}(\boldsymbol{\Psi}_h^n)\right), \qquad
\boldsymbol{\Psi}_h^{\ast\ast}
= \mathcal{E}_{\Delta t/2}
   \left(\boldsymbol{\Psi}_h^n,\,\mathcal{N}(\boldsymbol{\Psi}_h^\ast)\right), \qquad
\boldsymbol{\Psi}_h^{n+1}
= \mathcal{E}_{\Delta t}
   \left(\boldsymbol{\Psi}_h^n,\,\mathcal{N}(\boldsymbol{\Psi}_h^{\ast\ast})\right).
\end{align*}
On the large time step $\Delta t$, the slow advective transport of the velocity components $(u, v, w)$ and virtual potential temperature $\theta_v'$ is advanced using a third-order upwind-biased Eulerian flux scheme. For a scalar field $\chi$ transported in coordinate direction $\xi_d$, where $(\xi_1,\xi_2,\xi_3)=(x,y,\zeta)$, let $U^{(d)}_{i-1/2}$ denote the face-normal advecting velocity at the face between cells $i-1$ and $i$. Suppressing the transverse indices, the reconstructed face value is obtained by combining a fourth-order centred interpolation with a third-order upwind bias,
\begin{equation*}
\chi_{i-1/2} = \frac{7}{12}\left(\chi_i+\chi_{i-1}\right) -\frac{1}{12}\left(\chi_{i+1}+\chi_{i-2}\right) +\operatorname{sgn}\!\left(U^{(d)}_{i-1/2}\right)\frac{1}{12}\left[\left(\chi_{i+1}-\chi_{i-2}\right) - 3\left(\chi_i-\chi_{i-1}\right) \right] \,.
\end{equation*}
Near stagnation points, where $\lvert U^{(d)}_{i-1/2}\rvert<\epsilon_\ell$, the sign factor in the reconstruction is set to zero. This removes the upwind-bias term and recovers the fourth-order centred face value. The transported scalar flux, evaluated using the original, unmodified face-normal velocity, is
\begin{equation*}
F_{\chi,i-1/2}^{(d)} = U^{(d)}_{i-1/2}\chi_{i-1/2}.
\end{equation*} 
This suppresses sensitivity of the upwind-direction selection to round-off errors in nearly vanishing velocities. Away from this threshold, the third-order upwind formulation limits numerical dispersion while supplying controlled artificial diffusion to stabilize nonlinear advection.

Within each RK3 stage, let $\boldsymbol{\Psi}_{h,s}$ denote the fixed stage base state. The acoustic loop advances the fast correction
$
\boldsymbol{\delta\Psi}_h^{(\tau)} = \boldsymbol{\Psi}_h^{(\tau)} -\boldsymbol{\Psi}_{h,s},
$
whose velocity and pressure components are denoted by $\delta u$, $\delta v$, $\delta w$, and $\delta\pi$. The zonal momentum correction is advanced explicitly in a forward step, with an analogous equation for the meridional component, as
\begin{equation*}
\delta u^{(\tau+1)} = \delta u^{(\tau)} + \Delta\tau\left(\mathcal S_u -c_p\bar\theta_v m\frac{\partial\delta\pi^{(\tau)}}{\partial x} +D_{\mathrm{div},u} \right),
\end{equation*}
where $D_{\mathrm{div}}$ represents horizontal divergence damping evaluated continuously within the inner acoustic loop over $\Delta \tau$ to suppress acoustic energy accumulation before it can contaminate the slow modes.

To bypass restrictive vertical Courant--Friedrichs--Lewy (CFL) limits associated with thin vertical layers and steep terrain, vertical momentum and Exner pressure are solved implicitly in a backward step. Eliminating vertical velocity yields a decoupled one-dimensional vertical Helmholtz equation for the pressure increment $\pi_{\mathrm{inc}}^{(\tau)} = \delta\pi^{(\tau+1)}-\delta\pi^{(\tau)}$ across each vertical grid column,
\begin{equation*}
\pi_{\mathrm{inc}}^{(\tau)} - C_\pi \frac{\partial}{\partial z} \left( \gamma \frac{\partial \pi_{\mathrm{inc}}^{(\tau)}}{\partial z} \right) = \pi_{\mathrm{exp}}^{(\tau)},\quad \textup{where}\quad \pi_{\mathrm{exp}}^{(\tau)} =
\Delta\tau\left(\mathcal S_\pi-C_\pi\left[\operatorname{div}_h^{(\tau)} + \operatorname{div}_{z,\mathrm{exp}}^{(\tau)}
\right]
\right).
\end{equation*}
Here, $\operatorname{div}_h^{(\tau)}$ and $\operatorname{div}_{z,\mathrm{exp}}^{(\tau)}$ denote the horizontal and explicitly predicted vertical flux-divergence contributions, respectively. The wave coefficient is $\gamma = (\alpha \Delta \tau)^2 c_p \bar{\theta}_v \bar{\rho} \bar{\theta}_v$, $\alpha = 0.55$ is the acoustic off-centering weight, and $\pi_{\mathrm{exp}}^{(\tau)}$ contains explicit step-wise contributions from three-dimensional divergence and slow forcings. This system is inverted directly using a fast tridiagonal solver. The fast pressure correction is then updated by adding \(\pi_{\mathrm{inc}}^{(\tau)}\), and the vertical velocity is corrected using the vertical gradient of this solved pressure increment. Following the completion of all RK3 stages, passive tracer specific masses $q_i$ are advected over $\Delta t$ using the shared FFSL scheme described in Section~\ref{sec:spatial_discretization}, lateral boundary relaxation is applied, and the equation of state is re-applied to ensure exact thermodynamic consistency.

\subsection{Numerical diffusion and stabilization ($\boldsymbol{D}$)}
\label{sec:diffusion}

The diffusive terms $\boldsymbol{D}=(D_u,D_v,D_w)$ and $D_{\theta_v}$ in Eqn.~\eqref{eq:governing_eqns} collect the optional spatial filters applied within the dynamical core. These filters control unresolved grid-scale variance and computational modes. Their coefficients are numerical stabilization parameters rather than molecular viscosity or diffusivity. State-dependent turbulent mixing supplied by a physical closure is instead included in $\boldsymbol{P}$ and $P_{\theta_v}$, as described in Section~\ref{sec:physics} below. For each prognostic field $f\in\{u,v,w,\theta_v\}$, the core provides an optional fourth-order horizontal filter, $
D_f^{(4)}=-\nu_4\nabla_h^4 f$ where $\nu_4$ is scaled by a dimensionless factor relative to the explicit one-dimensional stability limit, $\Delta x^4/(64\Delta t)$. Rather than evaluating these operations by differencing along terrain-following $\zeta$-surfaces, each constituent Laplacian is constructed from the metric gradient \eqref{eq:grad_transform} and divergence operators~\eqref{eq:div_transform}, expressing the filter consistently in physical Cartesian coordinates on the terrain-following grid.

The horizontal momentum tendencies may additionally include divergence damping,
\begin{equation*}
\boldsymbol{D}_h^{\mathrm{div}}
 = \nu_{\mathrm{div}}\nabla_h
   \left(\nabla_h\cdot\boldsymbol{u}_h\right),
\end{equation*}
which suppresses grid-scale gravity-wave and acoustic noise. The coefficient $\nu_{\mathrm{div}}$ is scaled by a dimensionless factor relative to the explicit one-dimensional stability limit, $\Delta x^2/(4\Delta t)$. The timing of this filter depends on the dynamical core. In the semi-implicit semi-Lagrangian core, it is applied once per large time step $\Delta t$ to the implicitly updated horizontal winds, before lateral boundary blending and diagnostic equation-of-state reconciliation. In the Split-Explicit core, it is evaluated within each acoustic substep $\Delta\tau$ to suppress grid-scale acoustic oscillations. Thus,
\begin{equation*}
\boldsymbol{D}
 = \boldsymbol{D}^{(4)}+\boldsymbol{D}_h^{\mathrm{div}},
\qquad
D_{\theta_v}=D_{\theta_v}^{(4)}.
\end{equation*}
These smoothing filters are omitted when their corresponding coefficients are set to zero.

\subsection{Compact differentiable physics package
($\boldsymbol{P},P_{\theta_v},M_i$)}
\label{sec:physics}

The terms $\boldsymbol{P}$, $P_{\theta_v}$, and $M_i$ in Eqn.~\eqref{eq:governing_eqns} represent non-dynamical process contributions supplied through the differentiable physics interface. Unlike the fixed-form filters collected in $\boldsymbol{D}$ and $D_{\theta_v}$, these contributions may depend on the evolving atmospheric state and represent physical processes such as turbulent mixing, surface exchange, phase changes, and precipitation.

The interface supports two forms of coupling. Continuous schemes return rates of change that contribute to $\boldsymbol{P}$, $P_{\theta_v}$, or $M_i$ during the dynamical integration. Instantaneous schemes instead update the atmospheric state after a dynamical time step and therefore enter the time-discrete model through operator splitting rather than as an explicitly evaluated right-hand-side tendency.

As an example of a continuous closure, the package contains a stability-modulated Smagorinsky--Lilly subgrid turbulence scheme based on the resolved strain rate~\citep{smagorinsky1963general,lilly1966representation}. Its diagnosed turbulent-diffusion tendencies contribute to $\boldsymbol{P}$ rather than $\boldsymbol{D}$. As an example of an instantaneous adjustment scheme, the physics suite includes the two-category Kessler warm-rain scheme~\citep{Kessler1969} that includes prognostic equations for the mass mixing ratio of cloud water and rain water, in addition to water vapour. The scheme represents saturation adjustment and latent heating, autoconversion of cloud water, accretion, rain evaporation, and sedimentation. Sedimentation is integrated using a density-weighted finite-volume flux with internal substepping to satisfy its vertical Courant constraint, and precipitation leaving the lowest model level is accumulated at the surface. Its JAX implementation permits precipitation sensitivities to be propagated through the coupled dynamics and microphysics, as demonstrated in Section~\ref{sec:squall}.

\subsection{Boundary conditions}\label{sec:lbc}

The vertical domain is bounded by an impermeable rigid lid. At the uppermost $w$-face, the physical and contravariant vertical velocities satisfy
\begin{equation*}
w(x,y,L_z)=0,
\qquad
\dot{\zeta}(x,y,L_z)=0.
\end{equation*}
This ensures that the vertical mass flux through the top boundary vanishes. These constraints are imposed both during each time-stepping stage and then also re-applied after lateral boundary relaxation. No external-state relaxation is applied at the upper boundary as the Rayleigh sponge defined in Eqn.~\eqref{eq:rayleigh} damps upward-propagating waves under the rigid lid.

Lateral boundaries are treated with a Davies relaxation sponge \citep{davies1976lateral}. The relaxation weight follows a $\cos^2$ profile over a zone of width $D$ grid cells adjacent to each lateral wall,
\begin{equation*}
W(d) = \cos^{2}\!\left(\frac{\pi}{2}\,\frac{d}{D}\right), \qquad 0 \leqslant d \leqslant D \,,
\end{equation*}
where $d$ is the normal distance from the domain boundary measured in grid cells. After each time step, the prognostic subset $\{u, v, \theta_v, \pi', q_{i}\}$ is blended with the externally supplied driving state $\psi^{\mathrm{ext}}$ (derived, for instance, from global reanalysis),
\begin{equation*}
\psi^{n+1} \leftarrow (1 - C)\, \psi^{n+1} + C\, \psi^{\mathrm{ext}} \,,
\end{equation*}
using a direction-dependent total blending weight,
\begin{equation*}
C = \max\!\big(\, W_{\mathrm{inflow}},\ \alpha_{\mathrm{out}}\, W_{\mathrm{outflow}} \,\big), \qquad \alpha_{\mathrm{out}} \ll 1 \,.
\end{equation*}
This directional scaling strongly relaxes inflow boundary points, $C = W_{\mathrm{inflow}}$, toward the external driving fields while only weakly relaxing outflow boundary points, $C = \alpha_{\mathrm{out}} W_{\text{outflow}}$ with $\alpha_{\text{out}} = 0.01$, reducing spurious wave reflections at the domain exit interfaces. 

The physical vertical velocity $w$ and the contravariant vertical velocity $\dot{\zeta}$ are excluded from the blended set $\{u,v,\theta_v,\pi',q_i\}$. Instead, their boundary values are diagnosed from the kinematic conditions above after the horizontal winds have been blended. This preserves impermeability at both vertical boundaries and the local terrain-following condition at the lower boundary.

Finally, to prevent subgrid parameterizations from spuriously counteracting the relaxation sponge inside the lateral boundary zone, the model derives an interior mask equal to $1 - C$. This mask is passed directly to the extensible physics suite, multiplying and zeroing out all parameterized physical tendencies (such as turbulence and boundary-layer drag) within the boundary relaxation zone before they are injected into the dynamical core.\\

\noindent
\emph{Initialization and time-dependent lateral forcing from ERA5}
\vspace{0.1cm}

As a first step towards using Su\^{e}tes as a regional model for downscaling reanalysis data, ERA5 data can be supplied at its boundaries. A preprocessing pipeline is included which retrieves a user-selected ERA5 subset from the Copernicus Climate Data Store~\citep{https://doi.org/10.1002/qj.3803} and converts the pressure-level and surface fields into the model's native projected, terrain-following, staggered-grid state. The pipeline supplies both the initial atmosphere and the time-dependent external states used by the lateral relaxation in Section~\ref{sec:lbc}. Model topography can be constructed from the GEBCO elevation dataset~\citep{GEBCO2026}. This workflow is used for the Wreckhouse wind experiment in Section~\ref{sec:wreckhouse_sensitivities}, which is initialized from ERA5 and driven by hourly ERA5 lateral fields throughout the forecast.

The conversion accounts for the differences between the reanalysis and model grids without requiring users to construct a balanced Su\^etes state manually. ERA5 winds are rotated into the regional projection and placed on their C-grid faces, while atmospheric columns are interpolated in geometric height onto the terrain-following levels. Thermodynamic fields are converted to the prognostic variables of the core and reconciled with its discrete hydrostatic relation and equation of state to limit initialization noise. Vertical boundary velocities are then diagnosed consistently with the terrain and rigid lid. The processed hourly states are cached and interpolated in time during integration.

\section{Numerical validation and experiments}
\label{Sec:Results}

This section presents a sequence of test cases, which serve both as a forward validation of the model and as a demonstration of a distinct differentiable component. The rising thermal bubble and the Sch\"ar mountain-wave test verify the two dynamical cores against established benchmarks and against one another. The remaining cases combine a forward simulation with a gradient calculation, showcasing the model's differentiability with respect to the bottom topography, the initial distribution of a passive tracer, the initial thermodynamic state, the atmospheric state at an intermediate forecast time, and the parameters of the vertical coordinate transformation.

Table~\ref{tab:experiments} gives the numerical configuration of every case, grouped by the domain and grid, the time integration, the boundary treatment, the explicit dissipation, and the background state and prognostic tracers. The thermodynamic constants are $p_0=10^5\,\mathrm{Pa}$, $R_d=287\,\mathrm{J\,kg^{-1}\,K^{-1}}$, $c_p=1004\,\mathrm{J\,kg^{-1}\,K^{-1}}$, and $c_{vd}=717\,\mathrm{J\,kg^{-1}\,K^{-1}}$ throughout, and density is diagnosed from Eqn.~\eqref{eq:eos} in every case. Two cases are reported in two columns, since the Sch\"ar mountain-wave test and the squall-line benchmark are each first integrated forward and then differentiated in a separate configuration. Several cases were integrated with both dynamical cores but are shown for one only, and the corresponding column then describes the solution that is displayed. Quantities specific to a single case, namely the initial state, the terrain, the objective functional, and any optimiser settings, are specified in the respective subsections.

\begin{table}[p]
\centering
\rotatebox{90}{%
\begin{minipage}{\textheight}
\footnotesize
\setlength{\tabcolsep}{4pt}
\renewcommand{\arraystretch}{1.12}
\caption{Model configuration for each result shown in Section~\ref{Sec:Results}. For experiments run with both cores, but only one is shown, the configuration is that of the displayed solution.}
\label{tab:experiments}
\begin{tabular}{L{2.5cm} L{2.1cm} L{2.1cm} L{2.1cm} L{2.1cm} L{2.1cm} L{2.1cm} L{2.1cm} L{2.1cm}}
\toprule
 & \textbf{Rising bubble}
 & \multicolumn{2}{@{}l}{\textbf{Sch\"ar mountain}}
 & \textbf{Tracer inversion}
 & \multicolumn{2}{@{}l}{\textbf{Squall line}}
 & \textbf{Wreckhouse}
 & \textbf{NEUVE} \\
\cmidrule(lr){3-4}\cmidrule(lr){6-7}
 & (\ref{sec:bubble}) & Forward (\ref{sec:schaer}) & Opt.\ terrain (\ref{sec:schaer})
 & (\ref{sec:diff:tracer}) & Forward (\ref{sec:squall}) & Sensitivity (\ref{sec:squall})
 & (\ref{sec:wreckhouse_sensitivities}) & (\ref{sec:neuve_physics}) \\
\midrule

\multicolumn{9}{@{}l}{\itshape Domain}\\[1pt]
Extent $L_x\!\times\!L_y\!\times\!L_z$ (km)
 & $10\!\times\!0.15\!\times\!10$ & $100\!\times\!1.5\!\times\!20$ & $150\!\times\!1.5\!\times\!20$
 & $25\!\times\!25\!\times\!5$ & $150\!\times\!100\!\times\!24$ & $150\!\times\!100\!\times\!24$
 & $400\!\times\!400\!\times\!14$ & $16\!\times\!6\!\times\!16$ \\
Cells $n_x\!\times\!n_y\!\times\!n_z$
 & $200\!\times\!3\!\times\!200$ & $200\!\times\!3\!\times\!50$ & $300\!\times\!3\!\times\!50$
 & $50\!\times\!50\!\times\!20$ & $480\!\times\!320\!\times\!96$ & $150\!\times\!100\!\times\!24$
 & $200\!\times\!200\!\times\!40$ & $32\!\times\!12\!\times\!16$ \\
$\Delta x=\Delta y$, $\Delta z$ (m)
 & 50, 50 & 500, 400 & 500, 400 & 500, 250 & 312.5, 250 & 1000, 1000
 & 2000, 350 & 500, 1000 \\
\addlinespace[3pt]

\multicolumn{9}{@{}l}{\itshape Time integration}\\[1pt]
Core shown
 & Both & Both & Split-Expl. & Split-Expl. & Split-Expl. & Split-Expl.
 & Split-Expl. & Split-Expl. \\
$\Delta t$ (s), substeps
 & 1, 12 & 2, 6 & 4, 6 & 2, 6 & 0.375, 2 & 0.5, 2
 & 5, 4 & 0.8, 20 \\
$t_{\rm end}$
 & 1000\,s & 7200\,s & 1800\,s & 1200\,s & 7200\,s
 & 6750\,s (window from 6600\,s) & 9\,h & 400\,s \\
\addlinespace[3pt]

\multicolumn{9}{@{}l}{\itshape Boundaries}\\[1pt]
Lateral zone (cells)
 & rigid & 8, $x$ & 10, $x$ & 6 & periodic $x$; 10, $y$
 & periodic $x$; 10, $y$ & 15 & 4, $x$ and $y$ \\
Sponge base, $\tau_{\max}$
 & 7.5\,km, 0.05\,s$^{-1}$ & 12\,km, 0.5\,s$^{-1}$ & 12\,km, 0.5\,s$^{-1}$ & 4\,km, 0.5\,s$^{-1}$
 & 18\,km, 0.05\,s$^{-1}$ & 18\,km, 0.05\,s$^{-1}$ & 9\,km, 3.0\,s$^{-1}$
 & 12\,km, 0.3\,s$^{-1}$ \\
\addlinespace[3pt]

\multicolumn{9}{@{}l}{\itshape Dissipation}\\[1pt]
$\nu_h$, $\nu_v$ (m$^2$\,s$^{-1}$)
 & 0, 0 & 0, 0 & 0, 0 & 0, 0 & 33.33, 33.33 & 33.33, 33.33 & ---, --- & 0, 0 \\
$c_h$, $c_{\rm div}$
 & 0, 0 & 0, 0 & 0, 0 & 0, 0 & 0, 0 & 0, 0
 & 0.05 fwd.; 0.20 adj. & 0, 0 \\
\addlinespace[3pt]

\multicolumn{9}{@{}l}{\itshape State, physics and tracers}\\[1pt]
Stratification $N$ (s$^{-1}$)
 & 0 & 0.01 & 0.01 & 0.01 & W--K sounding & W--K sounding
 & ERA5, 14 Feb 2025 & 0.02 ($N_{\rm ref}=0.01$; sweep 0.01--0.02) \\
Physics
 & --- & --- & --- & --- & Kessler & Kessler & --- & --- \\
Tracers
 & --- & --- & --- & 1 passive & $q_v,q_c,q_r$, $\nu=0$ & $q_v,q_c,q_r$, $\nu=0$
 & --- & --- \\
\bottomrule
\end{tabular}
\end{minipage}}
\end{table}

\subsection{Rising thermal bubble}\label{sec:bubble}

The rising thermal bubble is a canonical test of buoyancy-driven, non-hydrostatic flow. A warm, positively buoyant perturbation is introduced into a dry, neutrally stratified atmosphere initially at rest. As the perturbation rises, horizontal buoyancy gradients generate vorticity along its flanks, and the thermal develops a mushroom-shaped cap with trailing counter-rotating vortices. The test therefore exercises nonlinear thermodynamic coupling, vertical acceleration, advection, and the interaction between acoustic and buoyancy modes. In Appendix~\ref{app:convergence} we describe convergence tests for both cores based on this test case. \\

\noindent\textbf{Forward simulation.} We consider a quasi-two-dimensional form of this benchmark which is evaluated on the full three-dimensional model grid using three homogeneous cells in the $y$ direction. Thus, in the following the $y$-coordinate is suppressed from the notation below. The reference virtual potential temperature is constant at $\bar{\theta}_v=\theta_0=300\,\mathrm{K}$, and the corresponding hydrostatic background Exner pressure is
\begin{equation*}
\frac{\mathrm{d}\bar{\pi}}{\mathrm{d}z}=-\frac{g}{c_p\theta_0},\qquad
\bar{\pi}(z)=1-\frac{gz}{c_p\theta_0}.
\end{equation*}
The flow is initially at rest,
\begin{equation*}
  u=v=w=0,\qquad\dot{\zeta}=0,\qquad\pi=\bar{\pi},
\end{equation*}
and the virtual potential temperature contains a cosine-squared perturbation,
\begin{equation*}
\theta_v(x,z,0)=\theta_0+\theta_v'(x,z),\qquad
\theta_v'(x,z)=
    \begin{cases}
      \Delta\theta
      \cos^2\left(\dfrac{\pi r}{2R}\right), & r\leqslant R,\\[1ex]
      0, & r>R,
    \end{cases}
\end{equation*}
where $\Delta\theta=2\,\mathrm{K}$, $R=1500\,\mathrm{m}$, and $r(x,z)=\sqrt{(x-x_c)^2+(z-z_c)^2}$ with $(x_c,z_c)=(0,2\,\mathrm{km})$. Davies lateral relaxation is disabled because this is an isolated idealized benchmark without externally prescribed boundary forcing. The thermodynamic variables are initialized in pressure equilibrium, with the Exner pressure equal to its hydrostatic background value. The temperature anomaly therefore produces a local density deficit and positive buoyancy without an initial Exner-pressure perturbation.

\begin{figure}[!ht]
 \centering
 \includegraphics[width=0.75\linewidth]{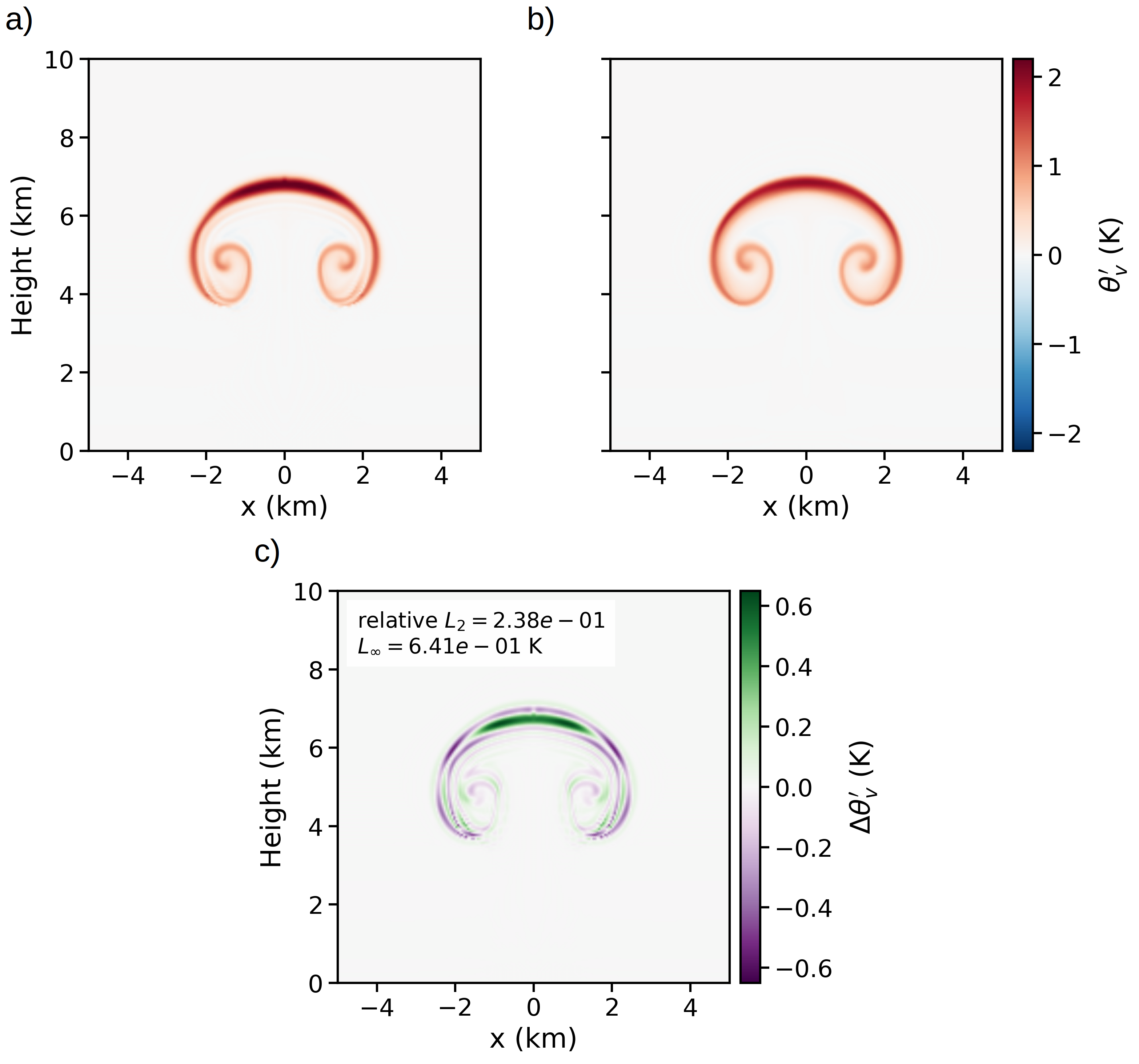}
    \caption{Virtual-potential-temperature perturbation on the central $x$--$z$ plane of the quasi-two-dimensional rising thermal bubble at $t=1000\,\mathrm{s}$ for (a) the SISL core, (b) the Split-Explicit core, and (c) their signed difference.}
    \label{fig:bubble}
\end{figure}

Figure~\ref{fig:bubble} shows that both cores produce the expected mushroom-shaped cap and the onset of two symmetric counter-rotating rotors. Their positive-anomaly centroids differ by only $5.3\,\mathrm{m}$, their vertical spreads by $3.0\,\mathrm{m}$, and their integrated positive temperature anomalies by approximately $0.5\%$. The cosine similarity between the complete perturbation fields is $0.9953$ and the relative $L_2$ difference is $9.69\%$. As apparent in Fig.~\ref{fig:bubble}(c), this difference is concentrated in narrow bands along the sharply varying thermal boundary and primarily reflects a small offset in interface position and amplitude, the peak-amplitude ratio being $1.142$, whereas the bulk position, spread, anomaly content, and rotor geometry remain closely matched. The left--right symmetry errors are below $2.5\cdot10^{-7}\,\mathrm{K}$ for both integrations and relative mass changes remain below $5.4\cdot10^{-7}$. Two independently formulated time-integration schemes therefore agree closely throughout the nonlinear development of the buoyant thermal.

\subsection{Schär mountain waves}\label{sec:schaer}

The Schär mountain-wave test~\citep{ANewTerrainFollowingVerticalCoordinateFormulationforAtmosphericPredictionModels} probes the treatment of steep, small-scale orography by the terrain-following coordinate. A uniform, stably stratified flow crosses a ridge whose Gaussian envelope excites vertically propagating gravity waves, while its shorter-scale modulation generates a trapped response near the surface. Since terrain-forced gravity waves govern mountain-wave amplitude, orographic precipitation, and severe downslope winds \citep{Durran1990, Kirshbaum2018}, the same configuration then serves as the forward model for an inverse problem, in which the bottom topography itself becomes the control variable. Both experiments in this subsection use a quasi-two-dimensional configuration evaluated with the full three-dimensional model state and numerical operators. The terrain and prescribed atmospheric fields are homogeneous across the three-cell $y$ direction, and the $y$-coordinate is suppressed from the notation below.\\

\noindent\textbf{Forward simulation.} The terrain is prescribed as
\begin{equation*}
    h(x)=h_0\exp\left[-\left(\frac{x}{a}\right)^2\right]
    \cos^2\left(\frac{\pi x}{\lambda}\right),
\end{equation*}
where $h_0=250\,\mathrm{m}$, $a=5\,\mathrm{km}$, and $\lambda=4\,\mathrm{km}$. The background virtual potential temperature and Exner pressure are
\begin{equation*}
 \bar{\theta}_v(z)=\theta_0\exp\left(\frac{N^2z}{g}\right), \qquad 
 \bar{\pi}(z)=1+\frac{g^2}{c_p\theta_0N^2}\left[\exp\left(-\frac{N^2z}{g}\right)-1\right],
\end{equation*}
with $\theta_0=300\,\mathrm{K}$, and satisfy $d\bar{\pi}/dz=-g/(c_p\bar{\theta}_v)$. The atmospheric fields are initialized as
\begin{equation*}
u=U,\qquad v=w=0,\qquad \pi=\bar{\pi},\qquad\theta_v=\bar{\theta}_v,
\end{equation*}
where $U=10\,\mathrm{m\,s^{-1}}$. At the first model step, the vertical boundary velocities are diagnosed consistently with the impermeability condition $\dot{\zeta}=0$. The lateral boundaries are relaxed toward this upstream state in the $x$ direction only, with $w$ and $\dot{\zeta}$ excluded so that the kinematic lower-boundary condition is preserved.

The initially uniform flow undergoes a short adjustment to the terrain boundary, producing a transient wave packet that is transported downstream. The stationary mountain-wave comparison is therefore evaluated over $|x|\leqslant20\,\mathrm{km}$, which excludes this startup transient and remains well separated from the lateral relaxation zones beginning near $|x|=46\,\mathrm{km}$.

\begin{figure}[!ht]
\centering
\includegraphics[width=\linewidth]{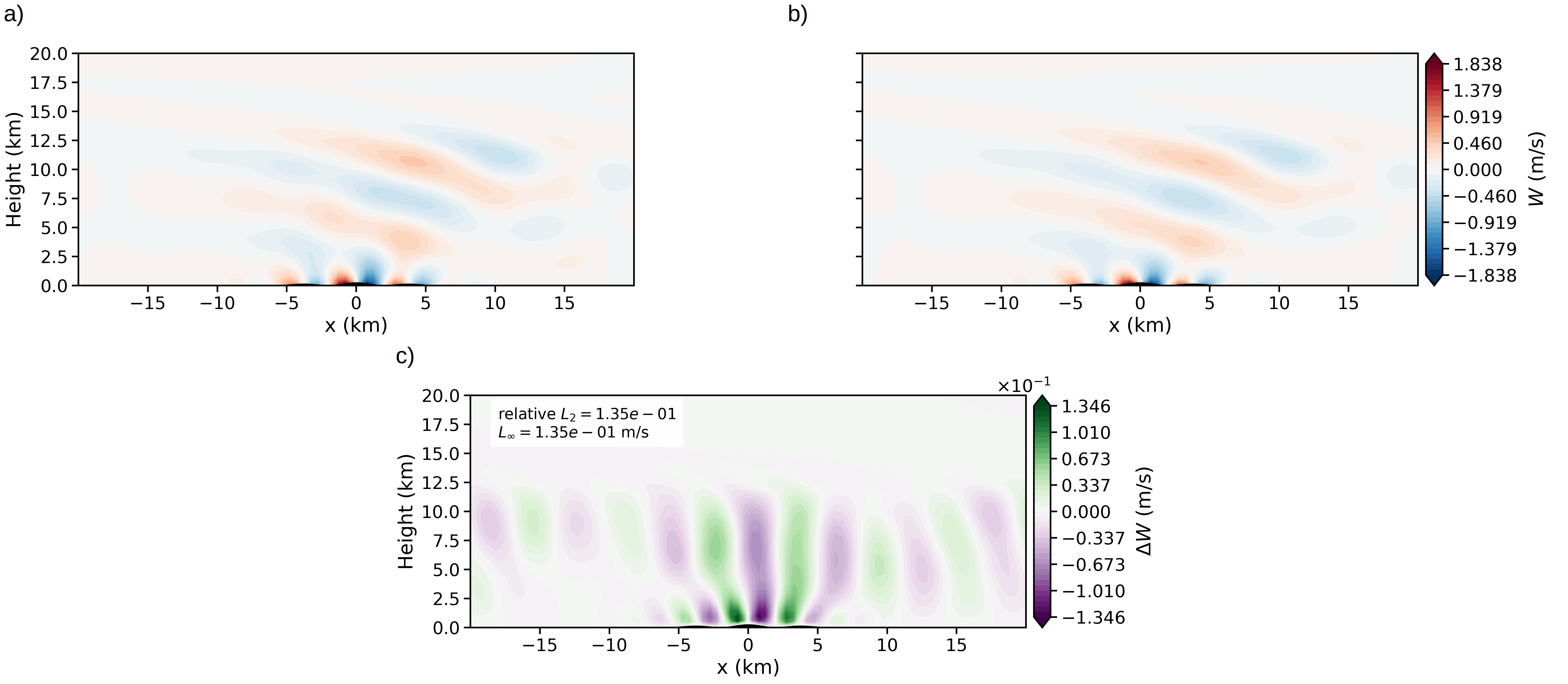}
\caption{Vertical velocity on the central $x$--$z$ plane of the quasi-two-dimensional Schär mountain-wave test at $t=7200\,\mathrm{s}$ for (a) the SISL core, (b) the Split-Explicit core, and (c) their difference. The displayed region, $|x|\leqslant20\,\mathrm{km}$, isolates the established terrain-forced wave from the downstream startup transient.}
\label{fig:schaer}
\end{figure}

Figure~\ref{fig:schaer} compares the vertical velocity produced by the SISL and Split-Explicit cores at the end of the integration. Both reproduce the terrain-following near-surface response and the broader vertically propagating wave generated by the Gaussian mountain envelope. Within the stationary-wave region the two solutions differ by $3.12\%$ in relative $L_2$ norm and have a cosine similarity of $0.999575$. Since the wave response is more usefully characterised by its vertical transport than by a pointwise difference, we also compare the \(x\)-integrated momentum flux per unit width in the homogeneous \(y\) direction, $M(z_k)=\sum_i \rho_{i,k}(u_{i,k}-U)w_{i,k}\,\Delta x$, evaluated with the staggered velocity components averaged to cell centres. These profiles differ by $0.89\%$ in relative $L_2$ norm and have a cosine similarity of $0.999973$.

\noindent\textbf{Gradient experiment.} We now optimize the terrain, under a fixed total mass, to maximize the gravity-wave energy in a chosen downstream region. The background flow and stratification are unchanged, on a domain extended downstream to accommodate the target region.

We use a dimensionally reduced representation of the terrain height to make the optimization more tractable by representing $h(x)$ as a linear combination of $N_{\rm rbf}=8$ Gaussian radial basis functions
\begin{equation*}
h(x) = \max\left(0, \sum_{i=1}^{8} A_i \exp\left(-\frac{(x - \mu_i)^2}{2\sigma^2}\right)\right),
\end{equation*}
where the standard deviation is fixed at $\sigma = 2.5$~km and the means $\mu_i$ are uniformly spaced between $x = -15$~km and $x = 10$~km. We impose a strict material budget to prevent the optimization from simply building an arbitrarily tall mountain by restricting the sum of the amplitudes to $A_{\rm tot} = 1500$~m, which fixes the cross-sectional area under the terrain for the shared width $\sigma$. Each $A_i$ is constrained by mapping unbounded optimization variables $z_i$ through a normalized sigmoid function as
\begin{equation*}
A_i = A_{\rm tot} \frac{\sigma(z_i)}{\sum_{j=1}^{8} \sigma(z_j)}, \quad i=1,\dots,8.
\end{equation*}
The objective is the sum of squared vertical velocity over the target region $x \in [15, 25]$~km, $z \in [3.2, 8.0]$~km, evaluated at the final time step,
\begin{equation*}
\mathcal{J} = \sum_{(i,k)\in\mathcal T}w_{i,k}(t_{\rm end})^2,
\end{equation*}
where $\mathcal T$ denotes the set of grid points in the target region. A reverse-mode pass through the integration yields the gradient $\nabla_z \mathcal{J}$ with respect to the eight topographic parameters. The terrain is optimized for 20 Adam iterations, with gradients clipped to a global norm of $1.0$ and a cosine-decaying learning rate from $0.5$ to $0.005$.

\begin{figure}[!ht]
    \centering
    \includegraphics[width=\linewidth]{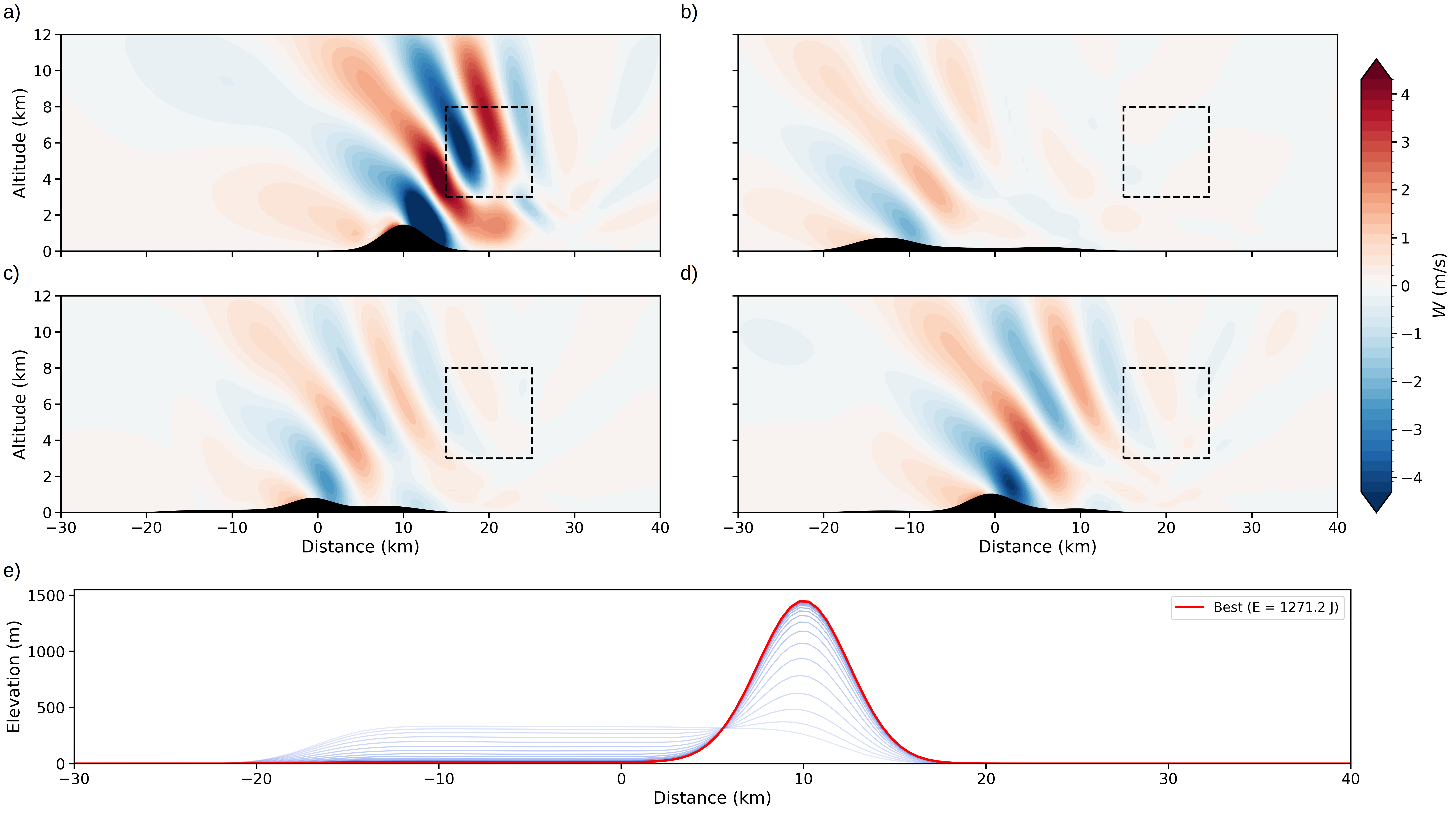}
    \caption{Vertical velocity at $t = 30$\,min for the optimal terrain (top left) and three random terrains drawn under the same total-mass budget. The dashed black rectangle marks the target box.}
    \label{fig:topo_validation}
\end{figure}

Figure~\ref{fig:topo_validation} compares the wave field generated by the optimal terrain against three random topographies drawn from the same mass budget. The optimal terrain uses its limited mass to phase the upward-propagating gravity waves so that they constructively interfere and peak within the downstream target box, delivering one to two orders of magnitude more target energy than the random mass distributions. Figure~\ref{fig:topo_validation}e traces the evolution of the topography throughout the optimization. Starting from a relatively flat, distributed initial guess, the optimizer systematically concentrates the available mass into a single steep ridge peaking near $x = 10$~km.

\subsection{Tracer source inversion over a complex terrain}\label{sec:diff:tracer}

In this test case, we consider a gradient-based optimization to recover the position and strength of an unknown tracer distribution from concentration measurements at a handful of downstream monitoring towers. Adjoint transport models are important tools for atmospheric tracer source inversion \citep{henze2007development}. Here, we consider an idealized passive-tracer experiment without chemical reactions to illustrate this capacity of the model. A Gaussian plume is initialized upstream of complex three-dimensional orography, in a flow with vertical shear in both wind components, so that the plume is tilted and spread as it is carried across the terrain and the tower records depend on the four source parameters in a strongly nonlinear way. Since the results are almost identical between the two cores, only the Split-Explicit solution is shown.\\

\noindent\textbf{Forward simulation.} The background flow is sheared and turning, with the westerly component increasing from $u_{\rm sfc}=5\,\mathrm{m\,s^{-1}}$ at the surface to $u_{\rm top}=15\,\mathrm{m\,s^{-1}}$ at the model lid, and a southerly component $v(z)=v_{\max}\sin(\pi z/H_{\rm top})$ with $v_{\max}=4\,\mathrm{m\,s^{-1}}$ and $H_{\rm top}=5\,\mathrm{km}$. The terrain carries two distinct peaks,
\begin{equation*}
h(x, y) = H_0 \exp\left( - \frac{(x - x_0)^2}{a_x^2} - \frac{(y - y_0)^2}{a_y^2} \right) + H_1 \exp\left( - \frac{(x - x_1)^2}{b_x^2} - \frac{(y - y_1)^2}{b_y^2} \right),
\end{equation*}
the primary peak reaching $H_0=1.8\,\mathrm{km}$ at $(x_0,y_0)=(-6,-2)\,\mathrm{km}$ with $a_x=a_y=5\,\mathrm{km}$, and the secondary peak $H_1=1.0\,\mathrm{km}$ at $(x_1,y_1)=(2,4)\,\mathrm{km}$ with $b_x=b_y=4\,\mathrm{km}$.

A passive tracer is initialized as a three-dimensional Gaussian plume,
\begin{equation*}
q_{\mathrm{tr}}(x, y, z;\, \mathbf{p}) = A \exp\!\left( -\frac{(x - x_s)^2}{2 \sigma_x^2} - \frac{(y - y_s)^2}{2 \sigma_y^2} - \frac{(z - z_s)^2}{2 \sigma_z^2} \right),
\end{equation*}
with fixed spreads $\sigma_x=\sigma_y=2\,\mathrm{km}$ and $\sigma_z=0.8\,\mathrm{km}$, and control parameters $\mathbf{p}=(x_s,y_s,z_s,A)$. The tracer is advected in flux form by the FFSL scheme alongside the dynamics.

\begin{figure}[!ht]
    \centering
    \includegraphics[width=0.95\linewidth]{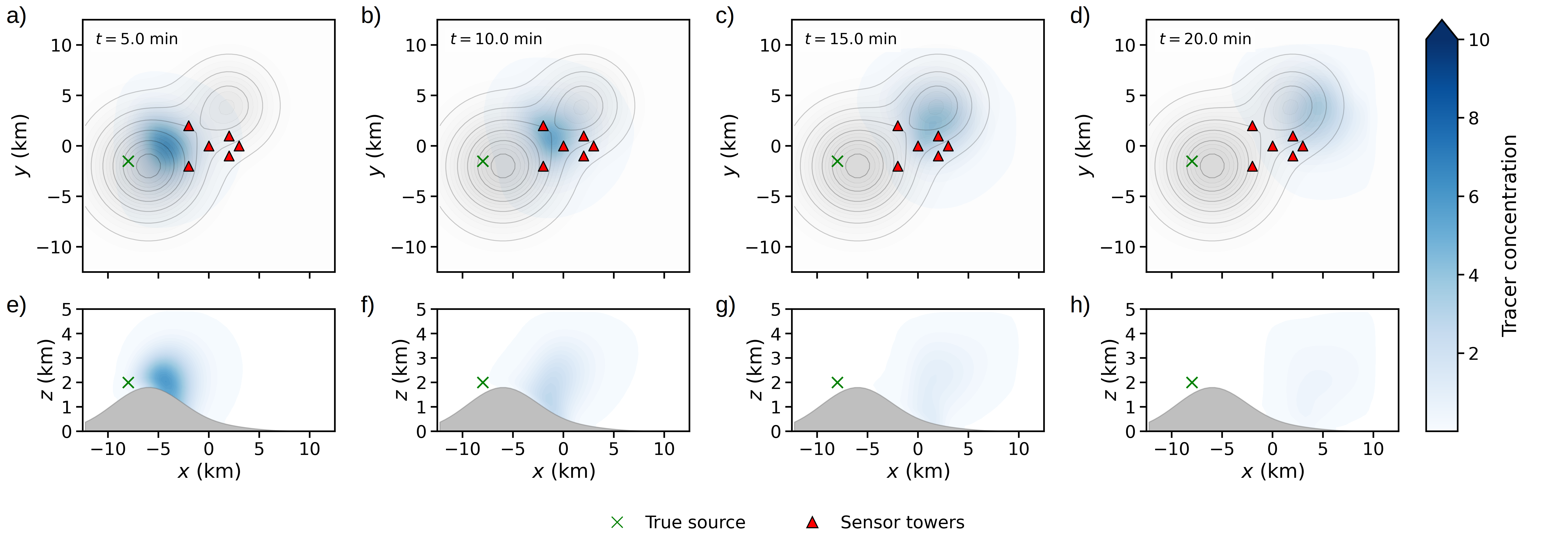}
    \caption{Temporal evolution of the true tracer plume over three-dimensional complex orography from $t=5$\,min to $t=20$\,min, showing horizontal $x$-$y$ slices (top row, $z \approx 2$\,km) and vertical $x$-$z$ slices (bottom row, along the $y = -1.5$\,km plane). The green cross marks the true source location.}
    \label{fig:tracer_snaps}
\end{figure}

Figure~\ref{fig:tracer_snaps} shows the plume produced by the true parameters $\mathbf{p}^{\star}=(-8\,\mathrm{km},\,-1.5\,\mathrm{km},\,2\,\mathrm{km},\,10)$, in horizontal sections at $z\approx2\,\mathrm{km}$ and vertical sections along $y=-1.5\,\mathrm{km}$. The plume is lifted over the primary peak and displaced laterally by the turning wind.

\noindent\textbf{Gradient experiment.} Synthetic observations are generated by integrating the true state forward and recording the tracer concentration after every time step at the grid cell nearest each of six monitoring towers,
\begin{equation*}
\mathbf{x}_s^{\rm obs} \in \Big\{ (-2, -2, 1), (-2, 2, 1), (0, 0, 1.5), (2, -1, 1.5), (2, 1, 2), (3, 0, 2) \Big\}\,\mathrm{km}.
\end{equation*}
The inverse problem minimizes the mean squared misfit between the simulated and target histories,
\begin{equation*}
\mathbf{p}^{\dagger} = \mathop{\arg\min}_{\mathbf{p}}\ \mathcal{J}(\mathbf{p}), \qquad \mathcal{J}(\mathbf{p}) = \frac{1}{N_s N_t} \sum_{s=1}^{N_s} \sum_{n=1}^{N_t} \Big( q_{\mathrm{tr}}\big(\mathbf{x}_s^{\rm obs}, t_n;\, \mathbf{p}\big) - \hat{q}_s(t_n) \Big)^{2},
\end{equation*}
where $N_s=6$ is the number of stations and $N_t$ the number of time steps in the integration. A related benchmark was recently proposed by \cite{cardoso2026solver} for a differentiable non-hydrostatic coastal hydrodynamical model.

A reverse-mode pass through the forward integration supplies $\nabla_{\mathbf{p}}\mathcal{J}$, the gradient passing through the transport scheme and the coupled dynamics into the analytic expression for the initial plume. Optimization uses Adam over 100 steps with a cosine schedule decaying from $0.25$ to two percent of that value. Each update is followed by clipping $x_s$ and $y_s$ to $[-20,20]\,\mathrm{km}$ and by imposing $z_s\geqslant0.1\,\mathrm{km}$ and $A\geqslant0$. The initial guess $\mathbf{p}^{(0)}=(-4\,\mathrm{km},\,3\,\mathrm{km},\,3.5\,\mathrm{km},\,2)$ places the source at the wrong horizontal position, $1.5\,\mathrm{km}$ above the true altitude, and at a fifth of the true amplitude.

\begin{figure}[!ht]
    \centering
    \includegraphics[width=0.95\linewidth]{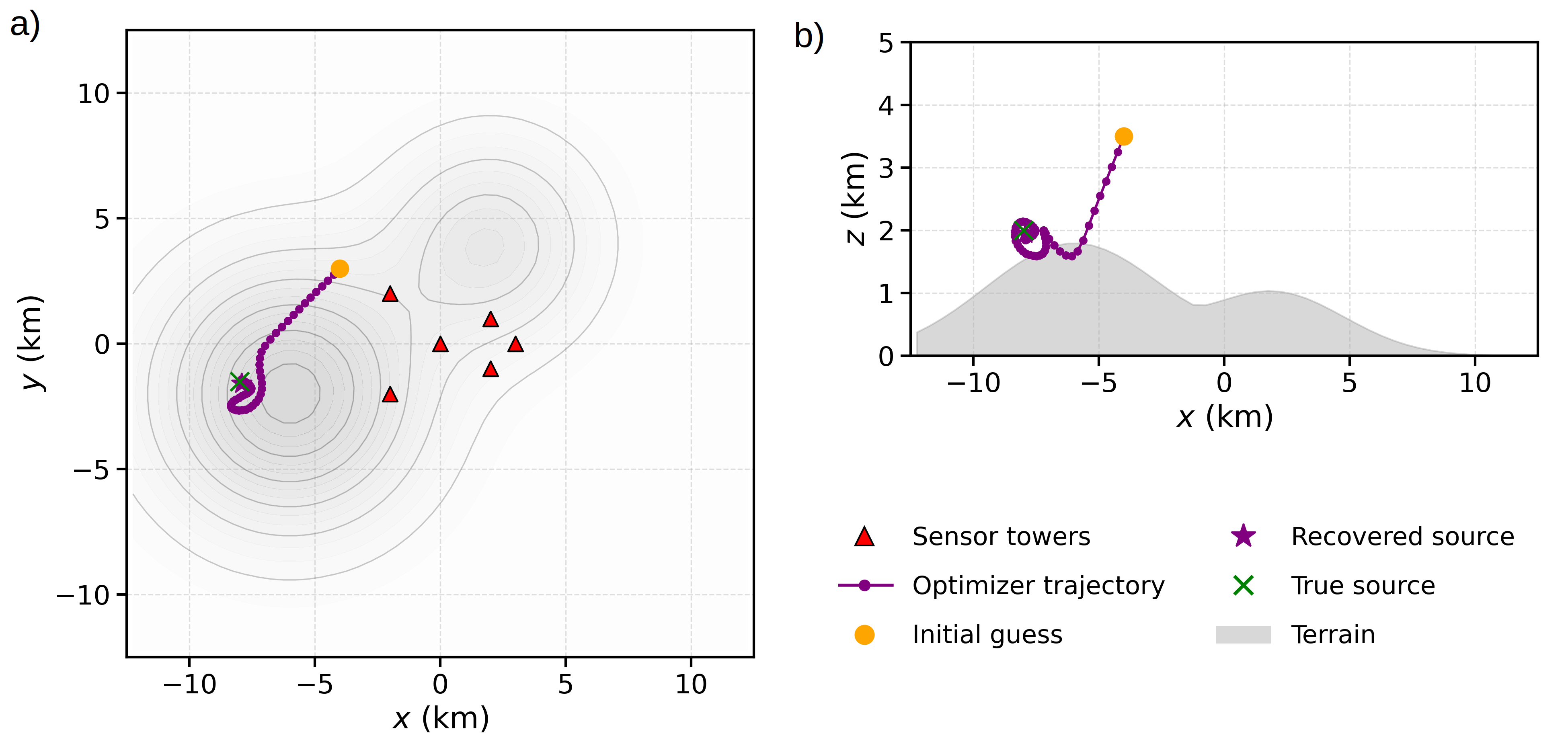}
    \caption{Optimizer trajectory in the horizontal $xy$-plane (left) and vertical $xz$-plane (right). The trajectory (purple) starts at the initial guess (orange) and converges to the true source (green cross) near the terrain profile (grey contours and silhouette).}
    \label{fig:tracer_inv}
\end{figure}

Figure~\ref{fig:tracer_inv} traces the optimizer in the horizontal and vertical planes. Starting from the wrong coordinates, it converges smoothly to the true source location within the 100 steps. 

\subsection{Three-dimensional squall-line benchmark}\label{sec:squall}

We simulate the moist Weisman--Klemp convective-storm benchmark~\citep{squall} in a configuration based on the Energy Research and Forecasting Model benchmark suite~\citep{https://doi.org/10.1029/2024MS004884}. Water vapour, cloud water, and rain water, denoted by $(q_v,q_c,q_r)$, are transported as mass-conserving tracers and coupled to the Kessler warm-rain microphysics scheme. The case exercises the coupling between the resolved dynamics and a threshold-dependent physical parameterization, and the gradient experiment that follows differentiates surface precipitation through both. \\

\noindent\textbf{Forward simulation.} The background atmosphere follows the Weisman--Klemp sounding. Potential temperature increases with height to the tropopause at $z_{\mathrm{tr}}=12\,\mathrm{km}$, above which it continues into an approximately isothermal stratosphere extending to the model lid,
\begin{equation*}
 \theta(z)=\begin{cases}\theta_0+(\theta_{\mathrm{tr}}-\theta_0)(z/z_{\mathrm{tr}})^{5/4}, & z\leqslant z_{\mathrm{tr}},\\[2pt]
 \theta_{\mathrm{tr}}\exp\!\left[\dfrac{g(z-z_{\mathrm{tr}})}{c_pT_{\mathrm{tr}}}\right], & z>z_{\mathrm{tr}},
\end{cases}
\end{equation*}
where $\theta_0=300\,\mathrm{K}$, $\theta_{\mathrm{tr}}=343\,\mathrm{K}$, and $T_{\mathrm{tr}}=213\,\mathrm{K}$. The relative-humidity profile is
\begin{equation*}
 \mathrm{RH}(z)=\begin{cases} 1-\dfrac{3}{4}(z/z_{\mathrm{tr}})^{5/4}, & z\leqslant z_{\mathrm{tr}},\\[2pt]
\dfrac{1}{4}, & z>z_{\mathrm{tr}},\end{cases}\qquad q_v(z)=\min\!\left[\mathrm{RH}(z)q_s(T,p),q_{v0}\right],
\end{equation*}
with $q_{v0}=14\,\mathrm{g\,kg^{-1}}$ and saturation mixing ratio $q_s$ evaluated using the Tetens formula. Exner pressure and moisture are iterated to obtain a hydrostatically balanced initial profile.

Convection is initiated using a spherical cosine-squared thermal,
\begin{equation*}
\theta'(\mathbf{x})=\begin{cases}\Delta\theta\cos^2\!\left(\dfrac{\pi r}{2}\right), & r\leqslant1,\\
0, & r>1,\end{cases}
\qquad r^2=\left(\frac{x-x_c}{x_r}\right)^2+\left(\frac{y-y_c}{y_r}\right)^2+\left(\frac{z-z_c}{z_r}\right)^2,
\end{equation*}
where $\Delta\theta=3\,\mathrm{K}$, $(x_c,y_c,z_c)=(0,0,2\,\mathrm{km})$, $x_r=y_r=10\,\mathrm{km}$, and $z_r=2\,\mathrm{km}$. The background zonal wind is
\begin{equation*}
 u(z)=u_s+\frac{\mathrm{d}u}{\mathrm{d}z}\min(z,z_s), \qquad u_s=-12\,\mathrm{m\,s^{-1}},\quad
    \frac{\mathrm{d}u}{\mathrm{d}z}=4.8\times10^{-3}\,\mathrm{s^{-1}}, \quad z_s=2.5\,\mathrm{km},
\end{equation*}
and therefore increases by $12\,\mathrm{m\,s^{-1}}$ over the lowest $2.5\,\mathrm{km}$ before becoming constant aloft. The Laplacian diffusion acts isotropically on momentum and virtual potential temperature and the moisture tracers are left undiffused so that their transport remains conservative through the flux-form semi-Lagrangian scheme.

\begin{figure}[!ht]
\centering
\includegraphics[width=\linewidth]{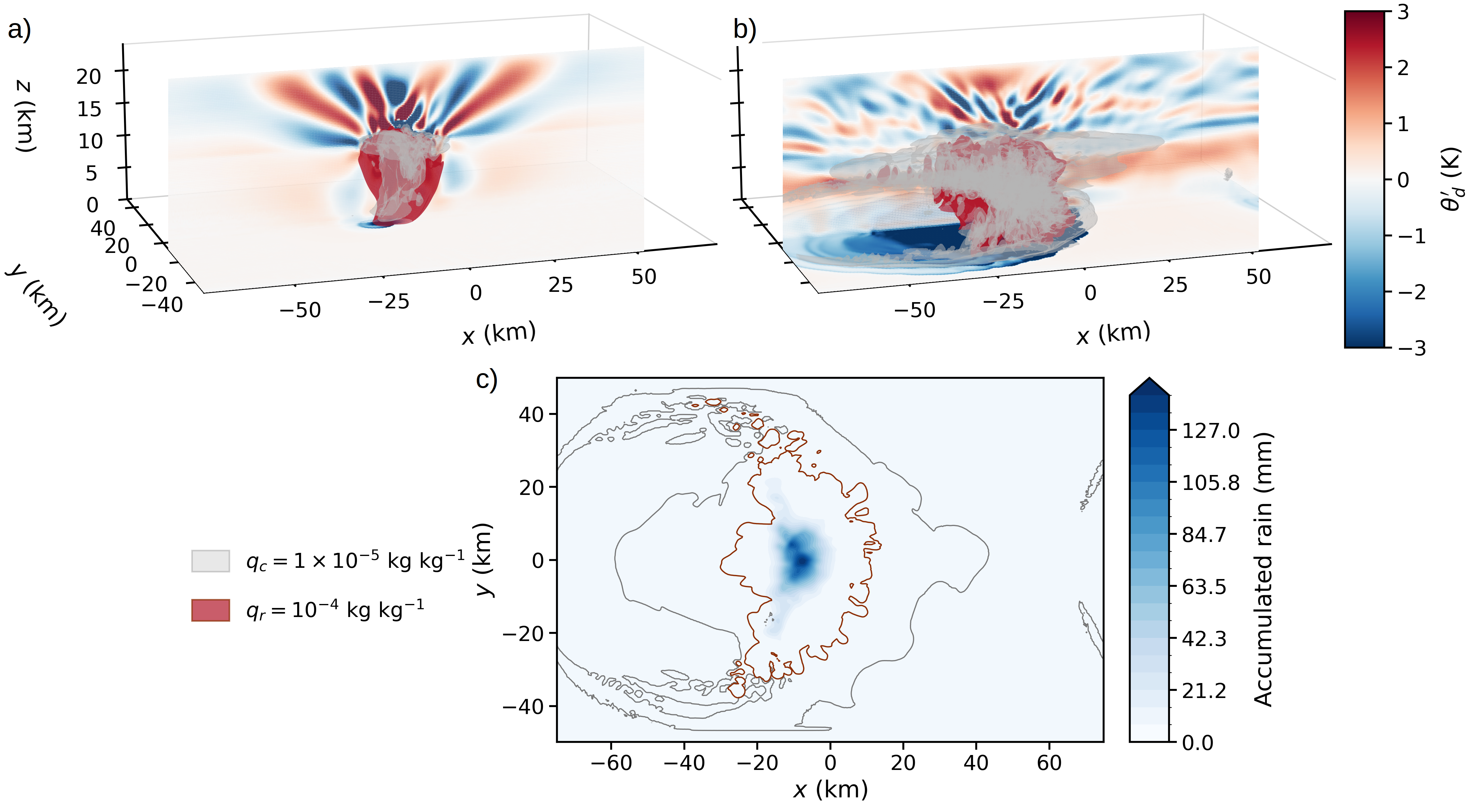}
\caption{Evolution of the three-dimensional Weisman--Klemp storm. Top: cloud-water and rain-water isosurfaces at $t=1800$ and $7200\,\mathrm{s}$, together with horizontal and vertical slices of the dry-potential-temperature perturbation $\theta_d'=\theta_d-\bar{\theta}_d(z)$, where $\bar{\theta}_d(z)$ is the initial horizontally homogeneous environmental sounding. Cloud water is shown at $q_c=10^{-5}\,\mathrm{kg\,kg^{-1}}$ and rain water at $q_r=10^{-4}\,\mathrm{kg\,kg^{-1}}$. Bottom: accumulated surface precipitation at $t=7200\,\mathrm{s}$.}
\label{fig:3D_squall}
\end{figure}

Figure~\ref{fig:3D_squall} shows the transition from the initially compact convective cell to a mature, horizontally extensive storm. By the end of the integration the cloud shield has developed a broad anvil, while rain water remains concentrated in localized shafts beneath the principal convective cores. The surface accumulation forms an elongated, multi-peaked swath aligned with the storm motion, reaching a maximum of $141.1\,\mathrm{mm}$.\\

\noindent\textbf{Gradient experiment.} The precipitation sensitivity is computed from a separate integration of the same benchmark on a coarser grid, spun up from rest to $t = 6600\,\mathrm{s}$ to produce a comparable mature storm. The control variables are additive perturbations to the dry potential temperature and water-vapour mixing ratio of that state, applied throughout the three-dimensional domain but masked from the lateral relaxation zones. Virtual potential temperature and density are recomputed from the perturbed fields so that the initial state of the differentiated window remains thermodynamically consistent.

The objective is the mean rain rate over the subsequent $150\,\mathrm{s}$ within a fixed storm footprint,
\begin{equation*}
\mathcal{J} = \frac{3600}{t\,|\Omega_s|}\sum_{(i,j)\in\Omega_s}\sum_{n=1}^{N_t}P_{ij}^{\,n}, \qquad t=150\,\mathrm{s},
\end{equation*}
where $P_{ij}^{\,n}$ is the surface precipitation accumulated during time step $n$. The target region $\Omega_s$ is diagnosed once from the unperturbed trajectory using interior columns exceeding $10\%$ of the maximum window accumulation, expanded by a four-cell halo, and is subsequently held fixed during differentiation. This avoids differentiating through the selection of the target itself.

A single reverse-mode pass provides the sensitivity of $\mathcal{J}$ to both control fields throughout the domain. For visualization, the horizontal responses are integrated over the lowest $3\,\mathrm{km}$ and scaled to uniform perturbations of $0.1\,\mathrm{K}$ and $0.1\,\mathrm{g\,kg^{-1}}$, respectively.

\begin{figure}[!ht]
\centering
\includegraphics[width=\linewidth]{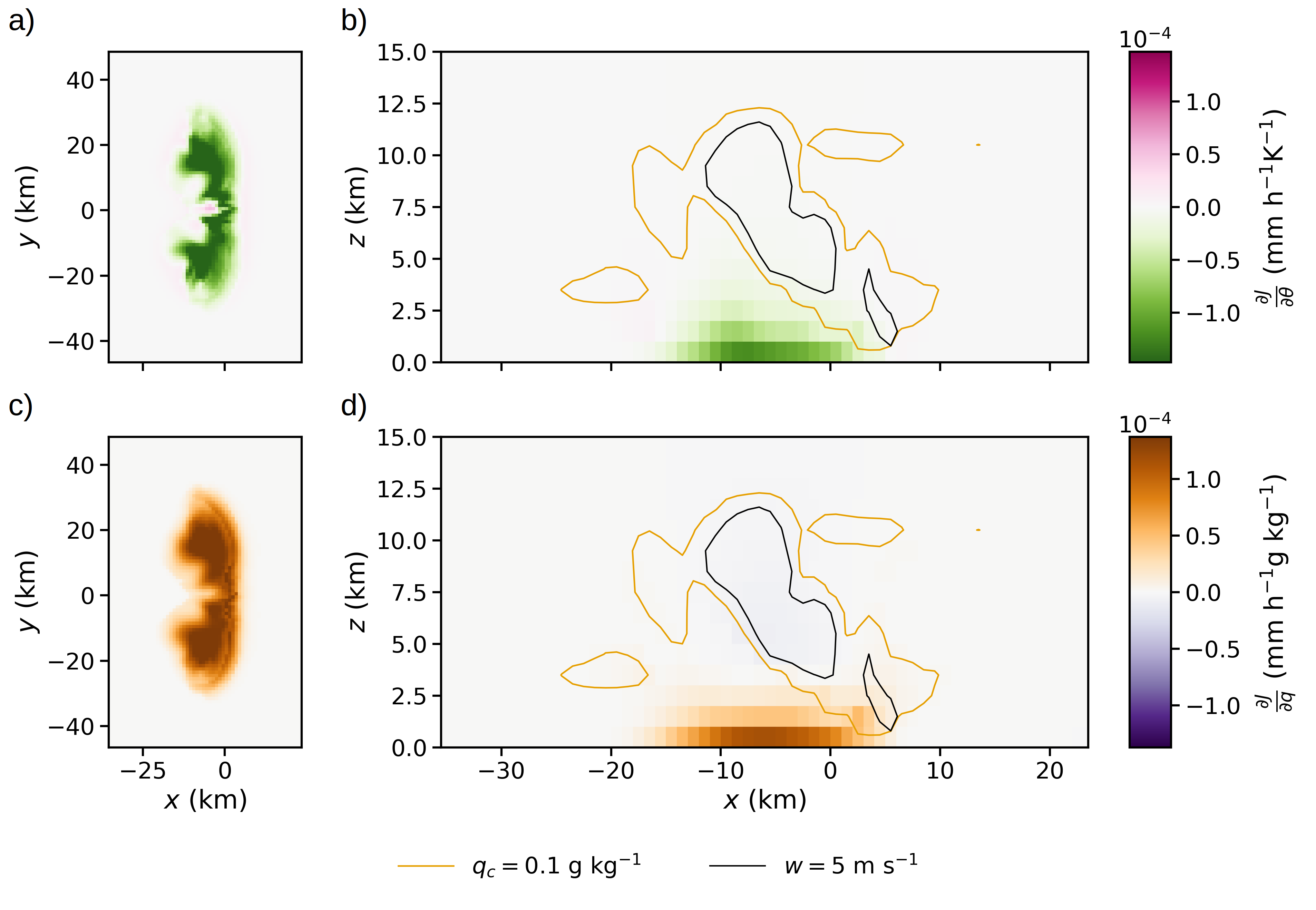}
\caption{Adjoint sensitivity of mature-storm precipitation. Top: accumulated precipitation over the target window and the linear rain-rate responses to boundary-layer perturbations in dry potential temperature and water vapour. The black contour denotes the fixed storm footprint and the dashed line marks the vertical-section location.  Bottom: sensitivities along a section through the strongest precipitating column. Orange and black contours show  $q_c=0.1\,\mathrm{g\,kg^{-1}}$ and $w=5\,\mathrm{m\,s^{-1}}$, respectively.}
\label{fig:squall_sensitivity}
\end{figure}

The sensitivities in Fig.~\ref{fig:squall_sensitivity} distinguish two thermodynamic controls on precipitation. Increasing low-level water vapour produces a coherent positive response across nearly the complete precipitation footprint, reflecting the increased condensate supply available to the mature updrafts. The temperature sensitivity is spatially signed and is predominantly negative beneath the active storm, indicating that a positive low-level dry-potential-temperature perturbation reduces subsequent precipitation to first order. This response is consistent with reduced relative humidity and condensation at fixed water-vapour mixing ratio, although the adjoint sensitivity combines these effects with changes in buoyancy, stability, and storm dynamics and does not isolate their individual contributions.

\subsection{The Wreckhouse winds}\label{sec:wreckhouse_sensitivities}

The \textit{Wreckhouse winds} of southwestern Newfoundland are severe downslope windstorms produced when strong flow crosses the Long Range Mountains and accelerates down the steep lee slope toward the coast. This case demonstrates differentiability in a real-world forecasting setting where an adjoint-derived sensitivity is used to construct a single localized thermal perturbation designed to increase the predicted wind at a specified verification time. \\

\noindent\textbf{Forward simulation.} The experiment is initialized and laterally forced by hourly ERA5 reanalysis fields \citep{https://doi.org/10.1002/qj.3803} for 14 February 2025, a date on which a Wreckhouse wind event occurred. The domain is centred on Wreckhouse, and its topography is derived from GEBCO elevations, lightly smoothed and blended into the lateral sponge. No subgrid physical parameterizations are active, so the response is produced by the resolved dynamics alone. The first six hours provide a dynamically adjusted baseline state.

\begin{figure}[!ht]
\centering
\includegraphics[width=\linewidth]{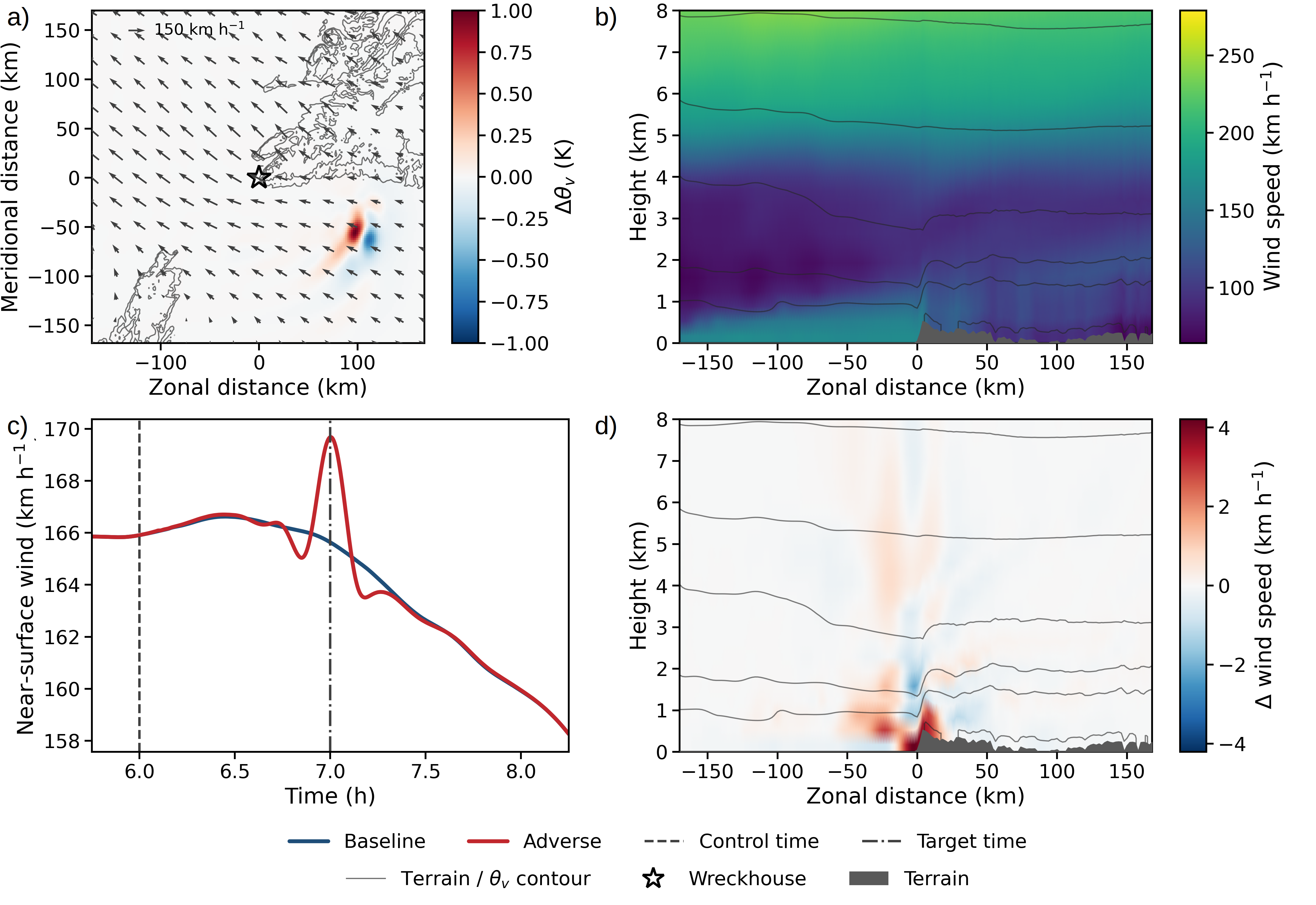}
\caption{Adjoint-directed modification of the Wreckhouse downslope-wind event. \textit{Top left:} Baseline and adverse wind speed at the central Wreckhouse grid point on the lowest terrain- following mass level, approximately 174~m above local terrain. The dotted and dashed vertical lines mark introduction of the thermal control at $t=6$~h and the verification time $t=7$~h, respectively. \textit{Top right:} Virtual potential-temperature perturbation on the lowest mass level at $t=6$~h, with terrain contours, baseline low-level winds, and the cross-section used in the lower panels, with the star denoting the Wreckhouse area. \textit{Bottom left:} Baseline wind speed and virtual-potential-temperature contours through the escarpment at $t=7$~h. \textit{Bottom right:} Adverse-minus-baseline wind-speed response at the same time, with baseline isentropes shown for context.}
\label{fig:wreckhouse}
\end{figure}

The baseline forecast contains a pronounced low-level jet with a wind speed maximum adjacent to the escarpment, accompanied by strongly displaced isentropes over the terrain, as shown in the lower panels of Fig.~\ref{fig:wreckhouse}.

\noindent\textbf{Gradient experiment.} At $t=6$~h the virtual-potential-temperature field is perturbed by an additive control $\Delta\theta_v$, distinct from the reference-state perturbation $\theta_v'$ of Section~\ref{sec:DynamicalCore}, and the resulting wind response is evaluated at $t_{\mathrm{opt}}=7$~h. Exner pressure is held fixed and density is re-diagnosed from Eqn.~\eqref{eq:eos} using $\theta_v+\Delta\theta_v$, so that the perturbed state remains thermodynamically consistent. No direct pressure or wind perturbation is applied.

The target diagnostic is the instantaneous spatial mean of the horizontal wind magnitude over a $3\times3$ grid-cell footprint centred on Wreckhouse, evaluated on the lowest terrain-following mass level, $k=0$. At the central target grid point this level lies approximately 219~m above sea level, or 174~m above the local terrain. Its geometric height varies horizontally with the terrain, and no interpolation to the surface or to a fixed height above ground is performed. After interpolating the staggered horizontal velocity components to mass points, the objective is
\begin{equation*}
\mathcal{J}(\Delta\theta_v)=\frac{1}{9}\sum_{(i,j)\in\mathcal{P}}\sqrt{u_m(i,j,k=0,t_{\mathrm{opt}})^2+v_m(i,j,k=0,t_{\mathrm{opt}})^2+\varepsilon},
\end{equation*}
where $\mathcal{P}$ denotes the target footprint and $\varepsilon$ regularizes the derivative at zero wind speed.

The control model used for differentiation retains the same resolved dynamics, grid, state at $t=6$~h, and lateral forcing as the forward evaluation, but applies stronger stabilization, $c_h=c_{\mathrm{div}}=0.20$ against $0.05$, to suppress grid-scale noise along the one-hour reverse trajectory. The resulting perturbation is then evaluated in the less strongly stabilized configuration, so the experiment tests whether a sensitivity obtained from the stabilized control model produces the predicted response under the standard settings.

A reverse-mode evaluation supplies the control-model gradient $g=\nabla_{\Delta\theta_v}\mathcal{J}$. To suppress grid-scale controls, $g$ is passed through a compact three-dimensional averaging operator $\mathcal{S}$ spanning $6\,\mathrm{km}\times6\,\mathrm{km}\times1.05\,\mathrm{km}$ and masked from the lateral relaxation zones. The applied perturbation is
\begin{equation*}
 \Delta\theta_v = \Delta\theta_{\max}\frac{\mathcal{S}g}{\|\mathcal{S}g\|_\infty},\qquad \Delta\theta_{\max}=1\,\mathrm{K}.
\end{equation*}
A one-sided Taylor test with a $0.01\,\mathrm{K}$ perturbation in this normalized direction gives a relative residual of $2.92\%$, confirming that the control-model gradient predicts the sign and local magnitude of the nonlinear response.

Figure~\ref{fig:wreckhouse} shows that the adjoint identifies a compact, signed thermal dipole in the upstream maritime flow rather than a spatially uniform warming. Although its pointwise magnitude reaches $1\,\mathrm{K}$, its domain-wide root-mean-square amplitude is only $0.015\,\mathrm{K}$. Its position relative to the baseline low-level winds indicates that the perturbation modifies the upstream thermodynamic environment encountered by the flow before it reaches the Long Range Mountains. In the vertical section through the escarpment, Fig.~\ref{fig:wreckhouse}(d), the perturbed and baseline runs differ by adjacent regions of strengthened and weakened flow, with the target lying in a strengthened one.

In the control configuration, the unperturbed objective at $t=7$~h is $164.9\,\mathrm{km\,h^{-1}}$. By applying the adjoint-directed perturbation in the evaluation configuration, where the unperturbed $3\times3$ grid-point-mean wind is $166.5\,\mathrm{km\,h^{-1}}$, the perturbed simulation reaches $170.5\,\mathrm{km\,h^{-1}}$, an amplification of $4.0\,\mathrm{km\,h^{-1}}$ or approximately $2.4\%$. At the central target point the wind increases from $165.6$ to $169.7\,\mathrm{km\,h^{-1}}$. The sensitivity therefore retains its predicted sign and produces a spatially coherent response despite the difference in numerical stabilization. The modest amplification is consistent with the deliberately constrained perturbation, capped at $1\,\mathrm{K}$ pointwise with a root-mean-square magnitude of $0.015\,\mathrm{K}$ and no direct perturbation to pressure or wind.

\subsection{Vertical coordinate tuning}\label{sec:neuve_physics}

A differentiable dynamical core can optimize not only atmospheric states and physical parameters, but also components of its numerical discretization. We demonstrate this capability by embedding a terrain-conditioned extension of the neural vertical coordinate NEUVE introduced by \cite{whittaker2025learningverticalcoordinatesautomatic} in the Su\^etes Split-Explicit core. In the original work, NEUVE was trained against the analytically solvable advection of a cosine bell over variable bottom topography. Here, no prescribed target coordinate or exact solution is supplied. Instead, the coordinate parameters are optimized directly against spurious motion generated by the discretized governing equations in an atmosphere that should remain at rest.\\

\noindent\textbf{Forward simulation.} A hydrostatically balanced resting atmosphere is an exact solution of the continuous non-hydrostatic equations. In terrain-following coordinates, however, the horizontal pressure-gradient force contains two opposing contributions,
 \begin{equation*}
  -\frac{1}{\rho}\left.\nabla_h p\right|_{z} = -\frac{1}{\rho}\left[\left.\nabla_h p\right|_{\zeta} - \frac{\nabla_h z}{\partial z/\partial\zeta}\frac{\partial p}{\partial\zeta}\right],
\end{equation*}
which cancel for the continuous resting solution. The vertical pressure-gradient and buoyancy terms must likewise remain balanced. Their discrete approximations need not cancel exactly on sloping coordinate surfaces, and the resulting residual accelerations generate spurious motion. The kinetic energy produced from an initially resting state therefore provides a self-supervised measure of the aggregate discrete balance error.

The dry test atmosphere is initially in hydrostatic balance and at rest. The Split-Explicit core decomposes the full thermodynamic state as the sum of a fixed hydrostatic reference profile and a perturbation, and we deliberately give the reference a stratification $N_{\mathrm{ref}}$ half that of the test atmosphere $N_{\mathrm{bv}}$. The initialized potential-temperature and Exner-pressure perturbations are therefore non-zero, despite the full atmosphere being balanced and at rest. In the continuous equations, any consistently constructed hydrostatic atmosphere remains at rest regardless of the reference profile used to represent it. Were the test atmosphere also used as the reference, the thermodynamic perturbations would vanish and the experiment would test only the equilibrium built directly into the model reference state. The factor-of-two contrast therefore provides a demanding but stable test of discrete well-balancedness, and reflects the normal forecasting situation in which a single fixed one-dimensional reference profile cannot match every horizontally varying and evolving atmospheric column.\\

\begin{figure}[!ht]
 \centering
\includegraphics[width=\textwidth]{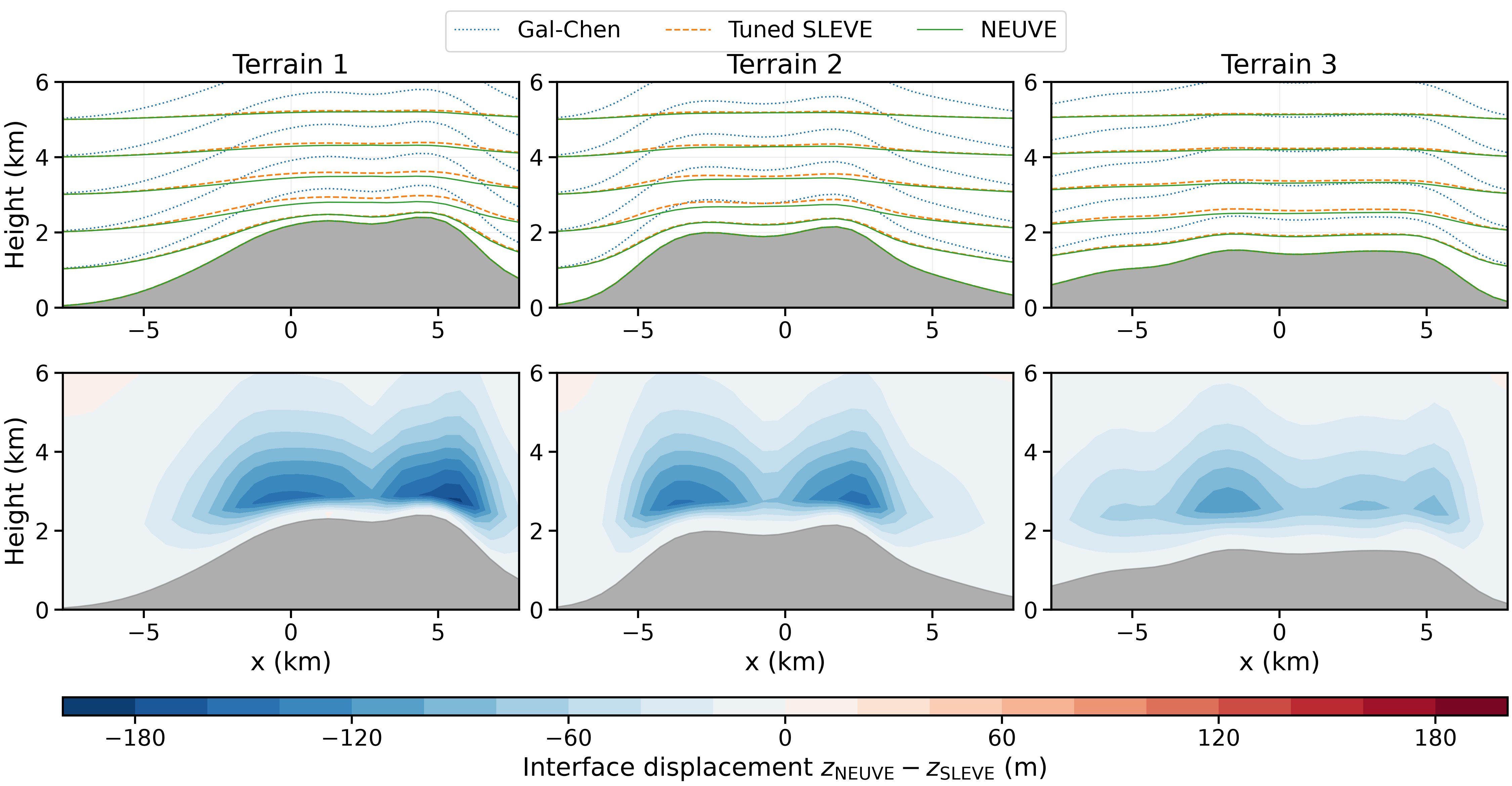}
\caption{Comparison of the Gal--Chen, training-set-tuned SLEVE, and NEUVE coordinates over three unseen terrains. The upper panels overlay coordinate interfaces within the lowest $6\,\mathrm{km}$, where terrain-induced deformation is concentrated. The lower panels depict the NEUVE--SLEVE interface displacement $z_{\mathrm{NEUVE}}-z_{\mathrm{SLEVE}}$. Both terrain-decaying coordinates flatten more rapidly than Gal--Chen and although their large-scale geometries are similar, NEUVE redistributes layer deformation locally in response to terrain structure.}
\label{fig:neuve_pgf_grids}
\end{figure}

\noindent\textbf{Gradient experiment.} The control variable is the vertical coordinate transformation itself, expressed through the terrain-decay function $B_\phi$ that sets how rapidly the influence of the topography diminishes with height,
\begin{equation*}
 z(\zeta,x,y)=\zeta+h(x,y)\,B_\phi(\eta,x,y),\qquad \eta=\zeta/L_z,
\end{equation*}
where $\eta$ is the normalized computational height. A valid coordinate requires $B_\phi$ to decrease monotonically from unity at the surface to zero at the lid. Rather than prescribing an analytical decay law satisfying these conditions, NEUVE builds $B_\phi$ as the normalized integral of a strictly positive layer density $\varrho_\phi$,
\begin{equation*}
 B_\phi(\eta,x,y)
 =\left(\displaystyle\int_0^1\varrho_\phi(s,x,y)\,\mathrm{d}s\right)^{-1}\displaystyle\int_\eta^1\varrho_\phi(s,x,y)\,\mathrm{d}s,
\end{equation*}
so that monotonicity, $B_\phi(0)=1$, and $B_\phi(1)=0$ hold by construction for any parameter values. Where $\varrho_\phi$ is large the coordinate surfaces bunch together, so the density governs the vertical distribution of layer compression. Its positivity is imposed by parameterizing the logarithm, which is unconstrained and is expanded in a clamped, open-uniform cubic B-spline basis $N_{k,3}$,
\begin{equation*}
 \varrho_\phi(\eta,x,y)=\exp\!\left[\ell_\phi(\eta,x,y)\right],\qquad
 \ell_\phi(\eta,x,y)=\sum_{k=1}^{K}c_k(x,y)\,N_{k,3}(\eta).
\end{equation*}
The basis is twice continuously differentiable and locally supported, giving a smooth, low-dimensional vertical profile, and the two integrals are evaluated with a fixed 12-point Gauss--Legendre rule, replacing the discrete cumulative integration of the original NEUVE formulation with an accurate, smooth evaluation at arbitrary computational heights.

The spline coefficients vary horizontally, combining a globally shared profile with a local correction conditioned on the terrain,
\begin{equation*}
 c_k(x,y)=\bar c_k+\alpha_s\,\delta c_k(\mathbf{q}(x,y);\psi),\qquad\mathbf{q} = \left(\frac{h}{h_{\rm s}}, |\nabla h|, h_{\rm s}\nabla_h^2h\right),
\end{equation*}
where $\bar c_k$ are trainable global coefficients and the terrain features $\mathbf{q}$ comprise the local height, slope, and curvature. The correction $\delta\mathbf{c}$ is produced by a multi-layer perceptron with weights $\psi$,
\begin{equation*}
 \delta c_k=\widetilde{\delta c}_k
   -\frac{1}{K}\sum_{j=1}^{K}\widetilde{\delta c}_j, \qquad  \widetilde{\delta\mathbf{c}} =\tanh\!\left(\mathbf{W}_2\mathbf{r}+\mathbf{b}_2\right),\quad
   \mathbf{r}=\tanh\!\left(\mathbf{W}_1\mathbf{q}+\mathbf{b}_1\right),
\end{equation*}
with $\psi=\{\mathbf{W}_1,\mathbf{b}_1,\mathbf{W}_2,\mathbf{b}_2\}$, $\mathbf{W}_1\in\mathbb{R}^{24\times3}$, and $\mathbf{W}_2\in\mathbb{R}^{K\times24}$. Subtracting the coefficient mean makes the conditioner redistribute the vertical decay rather than introduce an arbitrary common offset. The complete set of optimized coordinate parameters is therefore $\phi=\{\bar{\mathbf{c}},\psi\}$, and the terrain-conditioned extension can treat features of different geometry differently rather than applying one shared decay profile throughout the domain.

We set $K=16$, equal to the number of vertical model layers, giving the global profile approximately one locally supported degree of freedom per layer. This is a compact architectural choice rather than a hyperparameter selected against the held-out results. The terrain scale $h_{\rm s}=4000\,\mathrm{m}$ non-dimensionalizes height and curvature, and it keeps $h/h_{\rm s}<1$ for the generated topographies, whose elevations are capped below $2800\,\mathrm{m}$, while $h_{\rm s}\nabla_h^2h$ is dimensionless and numerically comparable to the other two features. The output layer of the conditioning network is initialized to zero, so the initial coordinate is determined entirely by the global spline profile, and the fixed factor $\alpha_s=1.5$ places the bounded corrections on the same order-one log-density scale as the global coefficients. These constants are held fixed for all training and testing terrains and are not selected using the held-out evaluation set.

The objective is the time-mean, volume-weighted spurious specific kinetic energy over the interior domain, subject to a barrier that keeps the layers from becoming arbitrarily thin,
\begin{equation*}
\phi^\dagger=\underset{\phi}{\arg\min}\;\mathcal{J}_{\mathrm{PGF}}(\phi),\qquad\mathcal{J}_{\mathrm{PGF}}(\phi)=\frac{1}{N_t}\sum_{n=1}^{N_t}
\frac{\displaystyle\sum_{i\in\Omega_{\mathrm{int}}}V_i\,\frac{1}{2}\left(u_i^2+v_i^2+w_i^2\right)}{\displaystyle\sum_{i\in\Omega_{\mathrm{int}}}V_i} + \mathcal{B}_{\Delta z}(\phi),
\end{equation*}
where
\begin{equation*}
\mathcal{B}_{\Delta z}(\phi) = \lambda_{\Delta z}\left[\max\left(0,\,\frac{\Delta z_{\min}-\min_i\Delta z_i(\phi)}{\Delta z_{\min}}\right)\right]^2, \qquad \Delta z_{\min}=150\,\mathrm{m},\qquad \lambda_{\Delta z}=100,
\end{equation*}
and $V_i$ is the cell volume. The barrier vanishes for mappings satisfying the minimum-layer constraint and penalizes thinner layers quadratically. Gradients are propagated through the complete integration, including the coordinate transformation and its metrics, the C-grid operators, and the Split-Explicit time integration.

NEUVE is trained jointly on ten independently generated three-dimensional terrains composed of randomly positioned, oriented, and anisotropic features over a range of horizontal scales. Before dynamical optimization, an inexpensive geometry-only bisection sets the global profile to the strongest uniform decay whose minimum training-set layer thickness remains above $175\,\mathrm{m}$, using neither SLEVE parameters nor dynamical-loss information. Starting from this mapping, the global spline coefficients and the conditioning network are optimized jointly for 60 epochs of Adam, with gradients clipped to a global norm of $1.0$, averaged across the ten terrains at each update, and driven by cosine-decaying learning rates of $10^{-3}$ and $3\cdot10^{-3}$ respectively. The reported coordinate is the checkpoint of lowest mean spurious kinetic energy.

For comparison, we tune a single-component analytical SLEVE coordinate on exactly the same ten-terrain training set,
\begin{equation*}
 z_{\mathrm{SLEVE}}=\zeta+h(x,y)
 \left[\frac{\sinh\!\left((L_z-\zeta)/s\right)}
 {\sinh\!\left(L_z/s\right)}\right]^n,
\end{equation*}
with two tunable parameters, the decay scale $s$ and the exponent $n$. It should be distinguished from the more general two-component SLEVE construction~\citep{ANewTerrainFollowingVerticalCoordinateFormulationforAtmosphericPredictionModels}, which first separates topography into large- and small-scale components and assigns them distinct decay profiles. The single-component form used here provides a well-defined two-parameter analytical comparator applied to the same undecomposed terrain field as NEUVE, and is subject to the same $150$-m minimum-layer constraint. A geometry search first determines the strongest admissible exponent for each of ten candidate decay scales, after which only the feasible boundary candidates are evaluated dynamically. The resulting SLEVE configuration is therefore a training-set-tuned analytical baseline rather than a default parameter choice.

\begin{figure}[!ht]
\centering
\includegraphics[width=\textwidth]{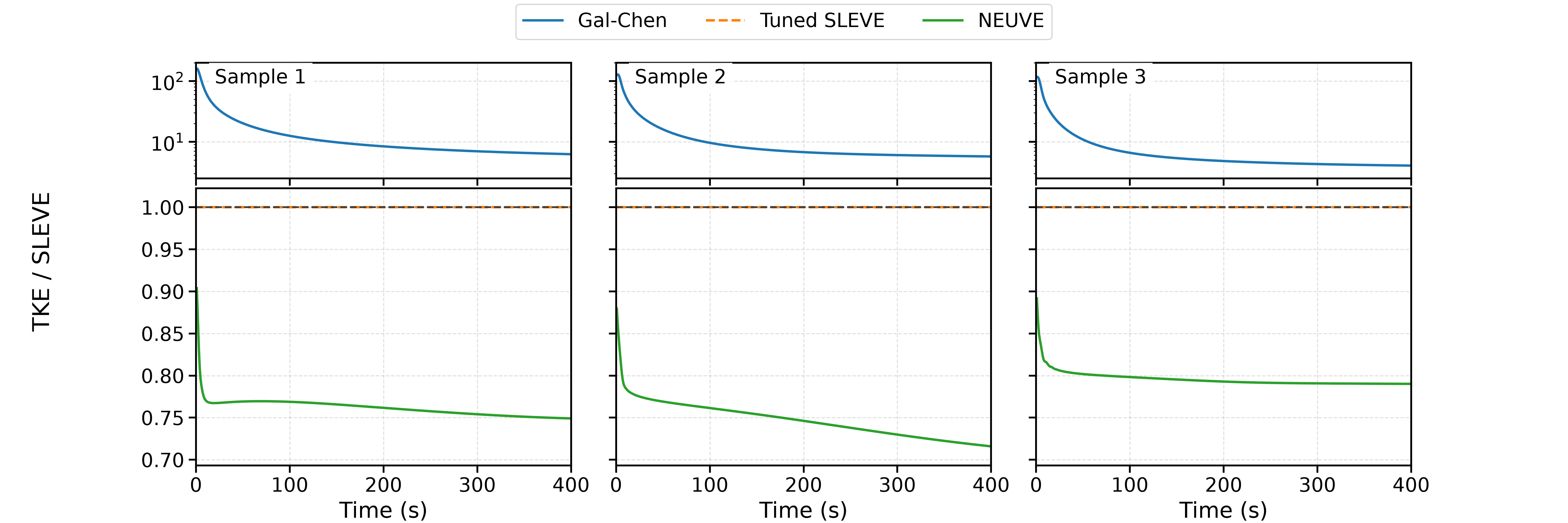}
\caption{Cumulative spurious specific kinetic energy relative to training-set-tuned SLEVE for three representative unseen terrains. The upper logarithmic panels show the much larger Gal--Chen response, while the lower panels resolve the difference between SLEVE and NEUVE. SLEVE defines the unit reference and values below unity indicate improvement over SLEVE.}
\label{fig:neuve_pgf_tke}
\end{figure}

Figure~\ref{fig:neuve_pgf_grids} situates both optimized terrain-decaying coordinates against Gal--Chen, which retains appreciably greater terrain deformation through the lower atmosphere. NEUVE does not obtain its advantage over tuned SLEVE from a radically different large-scale geometry: their interfaces remain close, but terrain-dependent displacements of up to approximately $200\,\mathrm{m}$ redistribute deformation and layer compression. The learned surfaces remain smooth while responding to changes in terrain height, slope, and curvature. The result indicates that minimizing the discrete pressure-gradient error depends not only on rapid upper-level flattening, but also on comparatively subtle changes in the distribution of slope, curvature, and layer compression throughout the column.

Figure~\ref{fig:neuve_pgf_tke} shows that Gal--Chen produces substantially more spurious motion than either terrain-decaying coordinate. NEUVE remains below the tuned-SLEVE reference throughout the integrations shown, rather than obtaining its advantage from an isolated final-time transient. For the three displayed terrains, the cumulative NEUVE-to-SLEVE ratio at the end of the integration is approximately $0.75$, $0.72$, and $0.79$. The magnitude of the improvement varies with terrain, demonstrating the value of adapting the coordinate to local topographic structure.

Across ten independently generated unseen terrains, Gal--Chen produces a mean spurious kinetic energy of $7.88\cdot10^{-1}\,\mathrm{m^2\,s^{-2}}$. Training-set-tuned SLEVE reduces this value to $1.23\cdot10^{-1}\,\mathrm{m^2\,s^{-2}}$, while NEUVE achieves the lowest error, $9.30\cdot10^{-2}\,\mathrm{m^2\,s^{-2}}$. Terrain by terrain, NEUVE correspondingly reduces spurious kinetic energy by $86.99\%$ relative to Gal--Chen and by $23.88\%$ relative to tuned SLEVE. The paired-bootstrap $95\%$ confidence interval for the improvement over SLEVE is $21.95$--$25.92\%$, and NEUVE outperforms SLEVE on all ten held-out terrains.

To determine whether this result depends on the particular stratification contrast used during training, we additionally evaluate the three fixed coordinates without retraining for hydrostatic resting states with $N_{\mathrm{bv}}\in\{0.01,0.0125,0.015,0.0175,0.02\}\,\mathrm{s}^{-1}$, retaining $N_{\mathrm{ref}}=0.01\,\mathrm{s}^{-1}$. In the matched-reference case, all coordinates produce mean kinetic energies of order $10^{-28}\,\mathrm{m^2\,s^{-2}}$, confirming that the model preserves its reference equilibrium to approximately machine precision, thus relative differences at this level are not meaningful. For the four off-reference atmospheres, the fixed NEUVE coordinate reduces spurious kinetic energy relative to fixed training-set-tuned SLEVE by $22.61$--$23.88\%$ and outperforms it on all ten terrains at every stratification. The spurious kinetic energy grows approximately quadratically with $N_{\mathrm{bv}}-N_{\mathrm{ref}}$ for all three coordinates. The learned advantage is therefore robust across the tested family of off-reference hydrostatic equilibria and is not specific to the factor-of-two contrast used during training.

Both optimized coordinates were constrained to a minimum layer thickness of $150\,\mathrm{m}$ on the training set. On the more extreme held-out terrains, the minimum thickness decreases to $81.0\,\mathrm{m}$ for tuned SLEVE and $78.7\,\mathrm{m}$ for NEUVE. All coordinate mappings remain monotonic and all integrations remain stable. Across the held-out set, NEUVE generally retains layer thicknesses comparable to SLEVE while producing smaller upper-level slopes and curvature. Its improvement is therefore not explained by permitting systematically thinner layers.

\section{Conclusions}
\label{sec:conclusions}

We have presented Su\^{e}tes, a fully compressible, non-hydrostatic, limited-area dynamical core written entirely in JAX. By reformulating the limited area model within a differentiable programming framework, Su\^{e}tes provides gradients of simulated atmospheric states with respect to topography, initial conditions, lateral boundary forcings, and physics parameters through a single reverse-mode pass. The dual-core architecture provides computational flexibility, offering both a SISL scheme for efficiency at coarser scales and a Split-Explicit Runge-Kutta scheme for high-resolution Eulerian dynamics. The core demonstrates robust performance across a suite of idealized benchmarks, including thermal bubbles, mountain waves and squall lines, as well as downscaling case studies driven at the lateral boundaries by state-of-the-art ERA5 reanalysis data.

The two time integrators are intended for complementary regimes rather than as interchangeable implementations with a universal performance ordering. The Split-Explicit core consists primarily of regular stencil operations and column-local solves. It is consequently well suited to accelerator execution and to convection-permitting simulations, where accuracy already requires relatively short advective time steps and where avoiding semi-Lagrangian interpolation helps retain short-wavelength structure. The SISL core instead trades a more costly step, including trajectory calculations and a Krylov solution of the coupled pressure--velocity system, for freedom from the advective and fast-wave time step restrictions. This trade is most attractive at coarser resolution or when a substantially longer outer time step can be used.

Reverse-mode differentiation introduces a second distinction. The Split-Explicit core uses check-pointing to limit storage of intermediate Runge--Kutta and acoustic-substep states, recomputing them during the reverse pass. The SISL core differentiates the implicit update through a transpose linear solve rather than through the stored sequence of forward Krylov iterations. The former favours regular computation, whereas the latter can reduce the storage associated with a long implicit solve. Which option is preferable therefore depends on resolution, attainable time step, solver convergence, rollout length, and available accelerator memory.

Our experiments confirmed these qualitative algorithmic differences, but timings of the present JAX implementation should not be interpreted as an intrinsic comparison of the two numerical methods. Kernel fusion, solver configuration, precision, hardware, and compilation strategy can all alter the balance. Appendix~\ref{app:performance} therefore reports only a compact CPU--GPU scaling experiment designed to demonstrate accelerator execution of representative forward and reverse workloads. Comparisons of the cores themselves are based primarily on their numerical properties and intended regimes of use.\par

The experiments performed in Section~\ref{Sec:Results} were designed to demonstrate and verify features of the model's differentiability. These test cases, however, represent idealized forms of problems of established scientific interest. Reconstructing an emission from downstream concentrations is the idealized form of source-term estimation, where the release history is unknown and the retrieval must be regularized against sparse and irregular observations \citep{henze2007development}. Identifying which parts of a ridge control the downstream wave response bears directly on the prediction of mountain waves, orographic precipitation, and severe downslope winds \citep{Durran1990, Kirshbaum2018}. The precipitation sensitivities point toward using gradients as a diagnostic of storm mechanism, asking which features of the environment a given storm's rainfall actually responds to rather than inferring this from ensembles of perturbed runs \citep{Derbyshire2004, SunCrook1997}. Perturbing an initial state allows one to construct consistent extreme-event storylines \citep{whittaker2025constructingextremeheatwavestorylines, hakim2026grayswanfactorymaking}. These applications are beyond the scope of this work, however, it is our hope that Su\^{e}tes' automatically differentiable dynamical core can provide a platform for extension of the model for these important problems.\par

Su\^{e}tes represents a first step toward a differentiable regional modeling system for numerical weather prediction and climate modeling applications. Hybrid architectures such as NeuralGCM~\citep{Kochkov2024} have shown what online-trained neural parameterizations can achieve globally, but the convection-permitting scales at which orographic flow, deep convection, and downslope windstorms organise have until now offered no differentiable core through which such components could be trained. A central accomplishment of this work is the integration of the model components into a common differentiable framework: terrain-following coordinate construction, spatial operators, advective transport, explicit and implicit time integration, lateral boundary relaxation, and the physical parameterizations included here all participate in the gradient calculation. This allows sensitivities to propagate through the simulated evolution to initial and boundary conditions, physical parameters, and the computational geometry itself, without separately maintaining a hand-written adjoint. 

Three developments would extend the system itself rather than its applications:\\

\noindent\emph{1) Large scale computations.} One natural extension is expanding the computational capability to distributed-memory architectures across multiple accelerators. The single-device experiments in Appendix~\ref{app:performance} corroborate that the present implementation already substantially benefits from modern hardware accelerators. At the largest tested grid, the GPU accelerates the Split-Explicit forward and differentiated workloads by a factor of approximately $60$ relative to the CPU, with corresponding factors of $38$ and $41$ for the SISL scheme. These limited hardware tests are not intended as optimized performance benchmarks, but they identify the distinct computational characteristics of the two cores. Split-Explicit consists mainly of regular stencil operations and column-local solves, whereas SISL additionally requires trajectory interpolation, Krylov iterations, and global reductions. Reverse-mode differentiation also increases memory use through stored or recomputed intermediate states as domain size and rollout length grow. Distributed domain decomposition, and multi-device checkpointing are therefore the next requirements for scaling Su\^{e}tes beyond a single accelerator. Such developments would enable larger regional domains, longer training rollouts, and high-resolution simulations of extended severe-weather events.\\

\noindent\emph{2) Differentiable physics packages.} A complementary direction is the development of a comprehensive, differentiable physics suite tailored for regional modeling. While Su\^{e}tes currently incorporates compact representations of basic moist processes and turbulence, a complete atmospheric model requires differentiable parameterizations of the boundary layer, radiative transfer, microphysics, and surface fluxes. Building on the methodology demonstrated by our neural vertical coordinate NEUVE~\citep{whittaker2025learningverticalcoordinatesautomatic}, learnable physics parameterizations can be constructed from scratch. Another option would be to calibrate against single-column frameworks~\citep{pierzyna2026jaxscmv10modernatmospheric} and high-resolution observational datasets. Developing these native closures---while drawing design principles from recent progress in other differentiable models, such as the JAX-based land-surface model~\citep{https://doi.org/10.1029/2024WR038116} and the JCM atmospheric physics package~\citep{egusphere-2025-6266}---will enable fully consistent, end-to-end gradient propagation through the coupled regional model. Another direction would be to follow NeuralGCM and develop a monolithic machine-learned physics parametrization.\\

\noindent\emph{3) Model coupling.} A third avenue lies in expanding towards multi-component Earth system modeling frameworks. By coupling Su\^{e}tes with external differentiable component models, the community can begin building fully coupled, end-to-end differentiable regional climate architectures. This could involve linking the atmospheric core to a differentiable ocean model, such as Veros~\citep{https://doi.org/10.1029/2021MS002717}, alongside hydrology and land-ice components. Furthermore, the high-resolution framework provides an ideal backbone for coupling with emerging differentiable wildland fire models~\citep{XIA2025106401, çakır2025jaxwildfiregpuacceleratedwildfiresimulator}, which leverage differentiable programming to optimize suppression strategies and simulator parameters via gradient descent. Coupling these processes would allow gradients to propagate across model interfaces, leading to a full picture of sensitivities across scales, from large-scale atmospheric drivers down to localized impacts.

\appendix
\section{Numerical verification of Su\^{e}tes}\label{app:convergence_verification}

This appendix provides supporting numerical information for the accuracy, differentiability, and computational performance of Su\^etes. Section~\ref{app:convergence} provides a concise numerical verification for both dynamical cores and their discrete adjoints, notably self- and cross-core convergence results. Section~\ref{app:performance} presents a short CPU--GPU scaling experiment for forward and reverse-mode execution of Su\^etes.

\subsection{Convergence of the forward and reverse methods}\label{app:convergence}

We use the method of manufactured solutions to test the consistency and accuracy of the dynamical cores. We recover approximately second-order spatial accuracy for the staggered momentum, Exner-pressure, and thermodynamic tendencies, including map and terrain-coordinate metric terms, with finest-grid rates between $1.93$ and $2.05$. Fixed-grid acoustic and internal-gravity-wave tests both recover second-order temporal convergence for the pressure--velocity and buoyancy--thermodynamic couplings of both cores. The shared flux-form semi-Lagrangian tracer transport is conservative and, as a distinct numerical component common to both cores, exhibits approximately first-order temporal convergence in a Gaussian bump transport test.

The iterative SISL solver was checked independently of the integrated refinement tests. These solver-verification experiments use 64-bit arithmetic so that round-off does not obscure the Krylov errors, which are measured against a reference transpose solve with a tolerance of $10^{-12}$. A manufactured right-hand side for the complete scaled pressure--velocity operator gives a forward relative residual of $1.53\cdot10^{-7}$. For the transpose solve used by the custom implicit adjoint, errors are measured relative to a reference gradient computed with a GMRES tolerance of $10^{-12}$ and an iteration limit of 100. At $\Delta t=10\,\mathrm{s}$, eight iterations reduce the relative gradient error to $5.58\cdot10^{-5}$, while 15 and 20 iterations reduce it to $1.86\cdot10^{-8}$ and $9.23\cdot10^{-9}$, respectively. The iteration requirement decreases rapidly with the time step: errors below $10^{-4}$ are obtained with four iterations at $\Delta t=5\,\mathrm{s}$, two at $2.5\,\mathrm{s}$, and one at $1\,\mathrm{s}$. This shows that the forward and transpose Krylov errors can be made smaller than the discretization errors, while also explaining why SISL cost depends strongly on time step, tolerance, and preconditioner quality.

All refinement tests use the centred value $\alpha=0.5$ to isolate the underlying discretization order. Operationally, Su\^{e}tes uses the standard configuration $\alpha=0.55$ to slightly damp computational modes. The overall forward convergence of the dual dynamical cores of Su\^{e}tes is illustrated in Fig.~\ref{fig:bubble_convergence} through a self-convergence test using the rising thermal in a $10\,\mathrm{km}\times10\,\mathrm{km}$ Cartesian $x$--$z$ domain with three grid cells in the homogeneous $y$ direction. The initial virtual-potential-temperature perturbation has an amplitude of $2\,\mathrm{K}$ and a cosine-squared radial profile of radius $1.5\,\mathrm{km}$ centred at $z=2\,\mathrm{km}$. Both cores are integrated to the common final time $t_{\rm end}=24\,\mathrm{s}$ on grids with $\Delta x=\Delta z=250$, $125$, $62.5$, and $31.25\,\mathrm{m}$. The large time step is refined proportionally from $1$ to $0.125\,\mathrm{s}$, and the Split-Explicit core uses 12 acoustic substeps throughout. Fine solutions are restricted to the next-coarser grid before evaluating differences over an interior subdomain, excluding a $1.5\,\mathrm{km}$ strip adjacent to the $x$ and $z$ boundaries to reduce the influence of numerical boundary treatment. The exclusion width is fixed in physical units across all resolutions, so the norms compare solutions over the same physical region. The reported rates therefore characterize interior convergence. Throughout Figs.~\ref{fig:bubble_convergence}--\ref{fig:adjoint_gradient_convergence}, the plotted discrete norm is the root-mean-square difference over the $N_\Omega$ included grid points, $\|\Delta q\|_2=[N_\Omega^{-1}\sum_{i\in\Omega}(\Delta q_i)^2]^{1/2}$. Both independently formulated cores approach second-order convergence for velocity, Exner pressure, and virtual potential temperature.

\setcounter{figure}{0}
\begin{figure}[!ht]
   \centering
   \includegraphics[width=\linewidth]{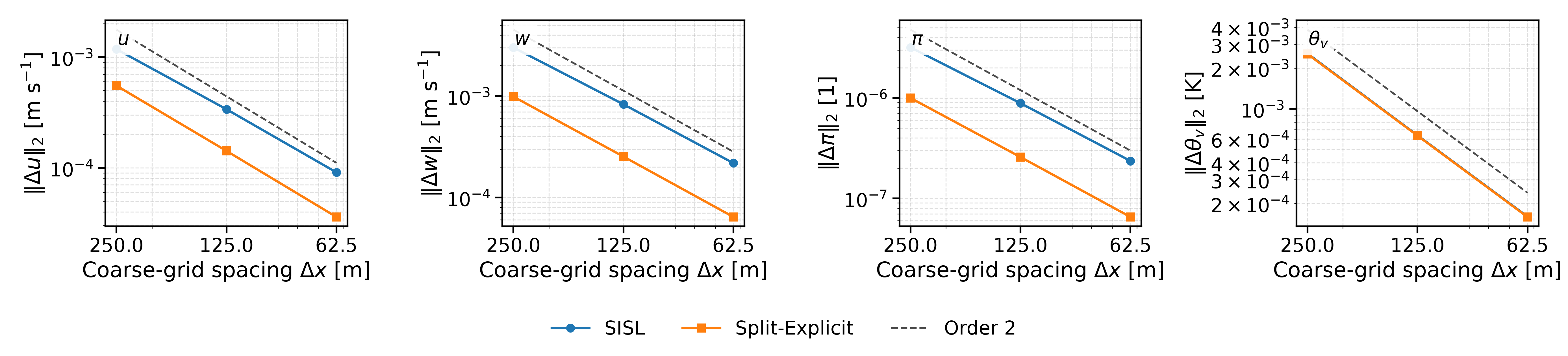}
   \caption{Combined space--time self-convergence of the Split-Explicit and SISL dynamical cores for the freely evolving rising thermal bubble. The panels show $L_2$ differences between successive restricted-grid solutions together with a second-order reference slope.}
   \label{fig:bubble_convergence}
\end{figure}

Since the two dynamical cores treat trajectories, acoustic modes, and pressure coupling fundamentally differently, their convergence to one another can serve as an additional core-health verification check. Figure~\ref{fig:cross_convergence} compares the SISL and Split-Explicit solutions from the same four integrations at $t=24\,\mathrm{s}$, using interior $L_2$ differences on each common grid. Their differences decrease systematically under refinement, with finest-refinement rates of $1.79$--$1.99$ across the prognostic fields. This supports the conclusion that both discretizations approach the same resolved solution rather than merely converging internally. 

\begin{figure}[!ht]
   \centering
   \includegraphics[width=\linewidth]{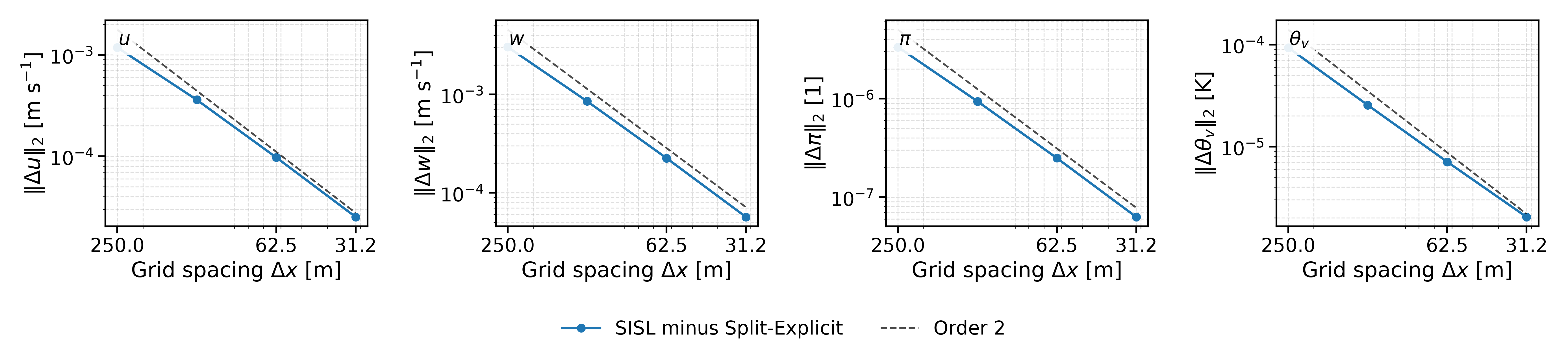}
   \caption{Cross-core convergence for the rising thermal bubble. The panels show the $L_2$ difference between the SISL and Split-Explicit solutions at each resolution together with a second-order reference slope.}
   \label{fig:cross_convergence}
\end{figure}

Reverse-mode derivatives are checked independently using a second rising-thermal experiment on a fixed $12\times3\times8$ grid with $\Delta x=\Delta z=375\,\mathrm{m}$ and final time $t=2\,\mathrm{s}$. The scalar objective combines volume-integrated updraft kinetic energy and virtual-potential-temperature perturbation variance. Its discrete gradient is evaluated with respect to the initial $u$, $w$, Exner-pressure perturbation, and cell-centred virtual potential temperature fields. Both cores use the centred, undamped refinement family with $\Delta t\in\{0.5,0.25,0.125,0.0625\}\,\mathrm{s}$.

\begin{figure}[!ht]
   \centering
   \includegraphics[width=\linewidth]{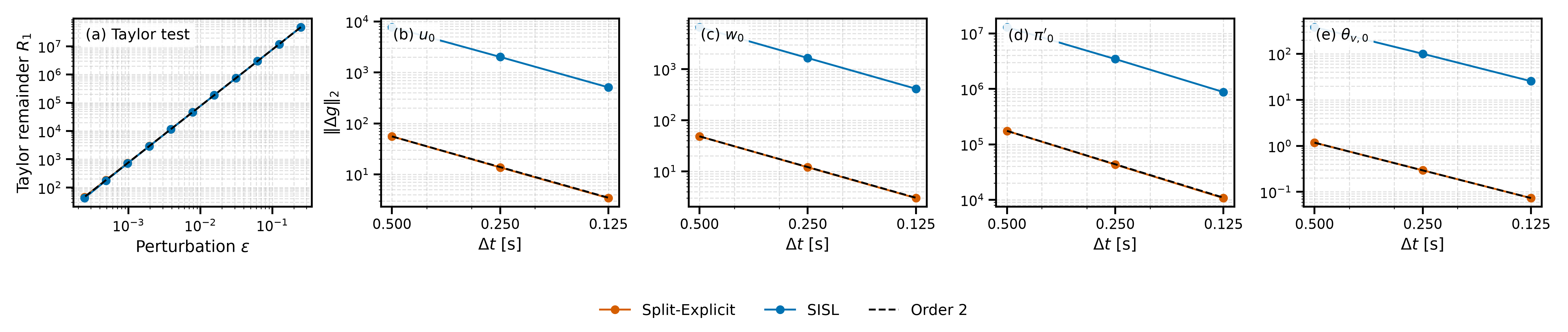}
   \caption{Verification and temporal convergence of the discrete adjoints. Panel~(a) shows the first-order directional Taylor remainder $R_1$ for perturbations to the initial virtual potential temperature. Panels~(b--e) show the RMS-normalized discrete norm $\|\Delta g\|_2$ of the difference between adjoint gradients at successive time steps for the initial $u$, $w$, $\pi'$, and $\theta_v$ fields, respectively. The large time step $\Delta t$ decreases to the right, while dashed lines denote second-order convergence.}
   \label{fig:adjoint_gradient_convergence}
\end{figure}

Figure~\ref{fig:adjoint_gradient_convergence}(a) shows quadratic decay of the first-order Taylor remainder after subtracting the adjoint directional derivative, directly verifying each discrete reverse-mode implementation. Panels~(b--e) show approximately second-order temporal self-convergence for all four gradient fields: the observed rates span $1.998$--$2.023$ for Split-Explicit and $1.911$--$2.008$ for SISL. For a separate $30\,\mathrm{s}$ nonlinear thermal experiment with a common objective, the independently generated virtual-potential-temperature sensitivity fields have a cosine similarity of $0.99955$. Thus the tests verify not only each discrete adjoint, but also temporal convergence across the prognostic controls and cross-core agreement in the resolved sensitivity structure.

\subsection{CPU--GPU execution scaling}\label{app:performance}

We use a limited hardware experiment to demonstrate accelerator execution of both JAX dynamical cores, not to establish an optimized performance baseline or rank the algorithms. Four-step rollouts are evaluated on an Intel Core i9-14900K CPU (32 logical processors) and an NVIDIA GeForce RTX 4090 GPU (24 GB). The dry, neutral $10\times10\times5\,\mathrm{km}$ domain is initially at rest apart from a spherical $2\,\mathrm{K}$ thermal of $1\,\mathrm{km}$ radius centred at $z=1.5\,\mathrm{km}$. Physics, lateral forcing, horizontal diffusion, and divergence damping are disabled. A weak $w$ sponge remains above $4\,\mathrm{km}$. Cell counts are refined as $N^3$, with $\Delta x=\Delta y=10\,\mathrm{km}/N$, $\Delta z=5\,\mathrm{km}/N$, and outer timesteps proportional to $\Delta x$. For each core and device, we time the compiled forward and complete value-and-gradient calculations. The scalar value combines the domain-mean final $w^2$ and virtual-potential-temperature perturbation variance, and the gradient is taken with respect to the complete floating-point initial state. Split-Explicit uses eight acoustic substeps and SISL uses a ten-times-longer outer timestep and GMRES with relative tolerance $10^{-4}$ and at most ten iterations. Compilation and warm-up are excluded, evaluations are synchronized, and repetitions start from the same state. Since the timestep changes under refinement and differs between cores, speed-ups are interpreted only within each core and grid size, thus absolute timings do not compare accuracy-normalized algorithms or resolutions. All timing experiments use 32-bit arithmetic on both devices.

\begin{figure}[!ht]
  \centering
  \includegraphics[width=0.7\linewidth]{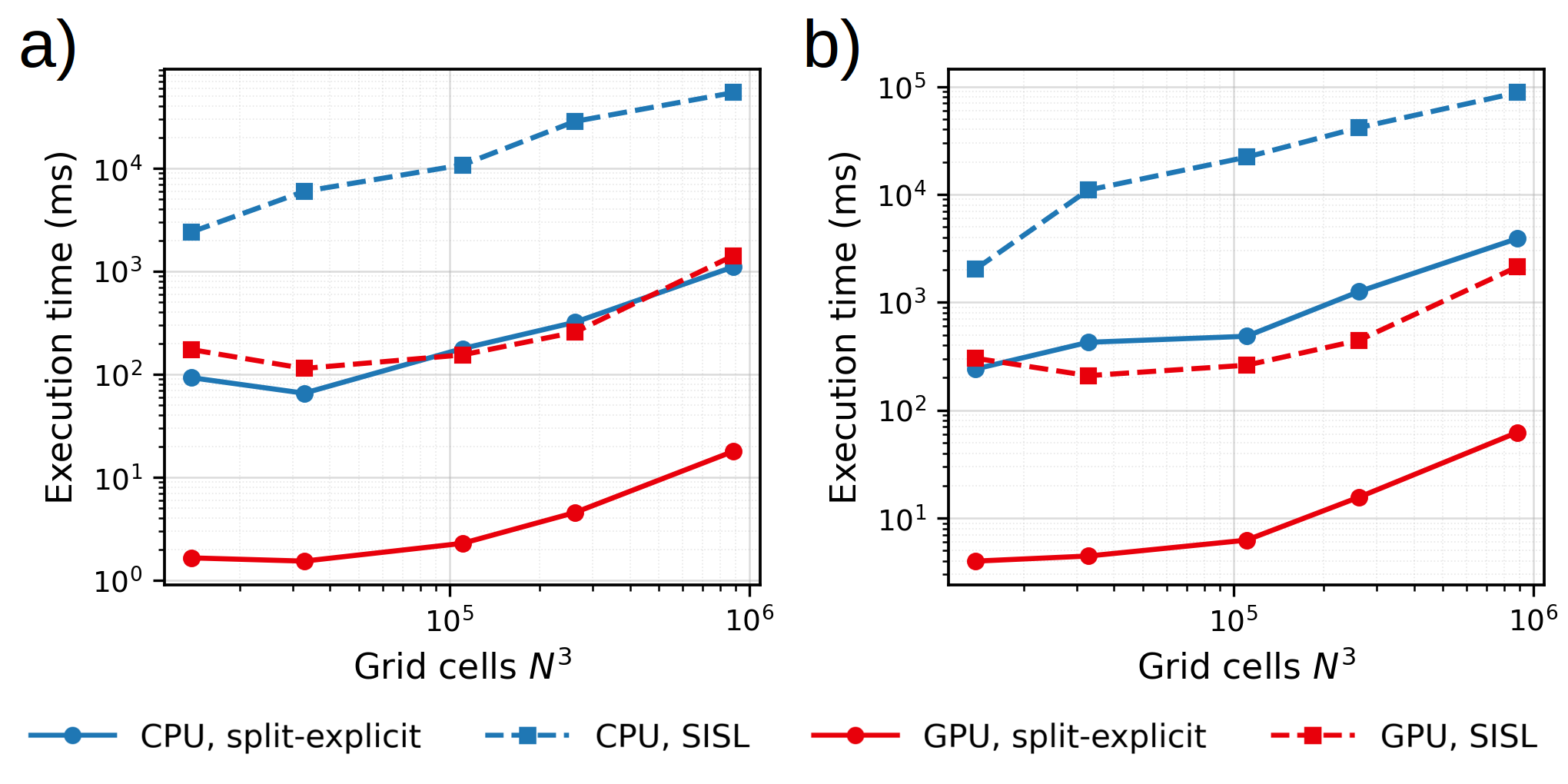}
  \caption{CPU--GPU execution scaling under three-dimensional grid refinement for short Split-Explicit and SISL rising-thermal rollouts on an Intel Core i9-14900K and NVIDIA GeForce RTX 4090. Panel~(a) shows compiled forward execution and panel~(b) the complete value-and-gradient evaluation, with both dynamical cores overlaid in each panel so the two workloads can be compared directly. Colour denotes platform (blue CPU, red GPU) and line style denotes core (solid circles Split-Explicit, dashed squares SISL). Timings exclude compilation and initialization and markers show medians over repeated synchronized evaluations.}
  \label{fig:cpu_gpu_scaling}
\end{figure}

The Split-Explicit measurements in Fig.~\ref{fig:cpu_gpu_scaling} show two execution regimes. At the smallest grids, GPU time changes little with problem size because launch latency and other fixed costs remain important. Throughput then increases rapidly as more parallel work becomes available. For the forward calculation it reaches $193$ and $230$ million cell-steps per second at $48^3$ and $64^3$ cells, respectively, before remaining of the same order at $96^3$. The value-and-gradient calculation reaches $71$ million cell-steps per second at $48^3$ and remains between $57$ and $67$ million cell-steps per second over the two larger grids.

In this large-grid regime, the GPU completes both Split-Explicit workloads approximately $60$ times faster than the CPU. At $96^3$ cells, four forward steps require $1.11\,\mathrm{s}$ on the CPU and $17.9\,\mathrm{ms}$ on the GPU, while the complete value-and-gradient calculation requires $3.90\,\mathrm{s}$ and $62.3\,\mathrm{ms}$, respectively. The complete differentiated calculation is therefore approximately $3.5$ times the cost of forward execution on either device, consistent with re-computation of check-pointed Runge--Kutta and acoustic-substep operations during the reverse pass.

SISL also benefits strongly from accelerator execution (Fig.~\ref{fig:cpu_gpu_scaling}), despite its less regular combination of trajectory interpolation, Krylov operator applications, and global reductions. Its GPU throughput increases through $64^3$ cells, reaching $4.08$ million cell-steps per second forward and $2.36$ million cell-steps per second for the complete value-and-gradient calculation, before decreasing modestly at $96^3$. At that largest grid, four SISL steps require $54.4\,\mathrm{s}$ forward and $87.9\,\mathrm{s}$ for value and gradient on the CPU, compared with $1.42\,\mathrm{s}$ and $2.13\,\mathrm{s}$ on the GPU. The corresponding accelerator speedups are $38$ and $41$, respectively. At $64^3$ cells they are larger, approximately $111$ and $94$, but we do not emphasize these peak ratios because CPU scaling and small-problem overhead vary across the refinement sequence.

The SISL value-and-gradient calculation at $96^3$ is only $1.5$ times its forward cost on the GPU and $1.6$ times on the CPU. This smaller relative reverse-mode overhead is consistent with differentiating the implicit update by solving the transpose system rather than retaining or replaying the forward GMRES iteration sequence. It does not make SISL universally faster: its forward step remains substantially more costly and its larger timestep advances a different physical duration. Rather, the Split-Explicit core maps efficiently to regular accelerator kernels, whereas SISL trades a more expensive implicit step for longer time steps and comparatively modest incremental adjoint cost.

These ratios are interpreted only as the speed-up achieved by this implementation for these representative workloads on the stated devices. They are not a comparison with other atmospheric models, a hardware-independent throughput claim, or an accuracy-normalized quantitative ranking of the SISL and Split-Explicit algorithms. Absolute timings remain sensitive to compiler and JAX versions, CPU threading, accelerator generation, precision, Krylov convergence, and problem size.

\clearpage		  
\acknowledgments
This work was supported in part by the Canada Research Chairs program and the Natural Sciences and Engineering Research Council of Canada (NSERC) Discovery Grant program. T.W.\ acknowledges support from an NSERC Canada Graduate Scholarship – Doctoral (CGS D) (Grant No. 588387-2024). S.T.\ acknowledges financial support from the Pacific Institute for the Mathematical Sciences and Environment and Climate Change Canada. The authors acknowledge the use of the large language models ChatGPT, Claude and Gemini to assist with software development. Any generated code was critically reviewed and revised by the authors, who take full responsibility for the scientific content, interpretations, and conclusions.

%
%
\availstatement
The code is publicly available on github (\url{http://github.com/Suetes/suetes}) with a static version for the paper on Zenodo (\url{https://doi.org/10.5281/zenodo.22667092}) \citep{bihlo_2026_22667092}. Most of the data was generated using Su\^{e}tes which is available at \url{http://github.com/Suetes/suetes}. ERA5 data was used to drive some of the cases and is openly available from the Copernicus Data Store (\href{https://doi.org/10.24381/cds.143582cf}{https://doi.org/10.24381/cds.143582cf}) \citep{https://doi.org/10.1002/qj.3803}

%






%



\clearpage

\bibliographystyle{ametsocV6}
\bibliography{references}

\end{document}